\documentclass[12pt,journal,draftclsnofoot,onecolumn,twoside]{IEEEtran}
\usepackage{indentfirst}
\usepackage{graphicx}
\usepackage{epstopdf}
\usepackage{amsfonts}
\usepackage{multirow}
\usepackage{amsmath,amsthm,amssymb}
\usepackage{color,changepage}
\usepackage{CJK}
\usepackage{algorithm}
\usepackage{algorithmic}
\usepackage{bbm}
\usepackage{verbatim}
\usepackage{makecell}
\usepackage{subfigure}
\usepackage{mathrsfs}
\usepackage{arydshln}
\usepackage{cases}
\usepackage{extarrows}
\usepackage{array}
\usepackage{bm}
\usepackage{amsmath}
\usepackage{setspace}
\allowdisplaybreaks[4]
\usepackage[T1]{fontenc}
\usepackage{threeparttable}
\usepackage{booktabs}
\usepackage{caption}
\definecolor{newcolor}{rgb}{0.5,0,1}

\newtheorem{Theorem}{Theorem}

\newtheorem{Lemma}{Lemma}

\theoremstyle{remark}
\newtheorem{Remark}{Remark}

\newcommand{\minitab}[2][l]{\begin{tabular}{#1}#2\end{tabular}}

\DeclareMathAlphabet{\mathpzc}{OT1}{pzc}{m}{it}

\definecolor{newcolor}{rgb}{0.5,0,1}

\begin{document}

\title{Improved Constructions for Secure Multi-Party Batch Matrix Multiplication}
\author{Jinbao Zhu, Qifa Yan, and Xiaohu Tang
\thanks{
The authors are with the Information Security and National Computing Grid Laboratory, Southwest Jiaotong University, Chengdu 611756, China (email: jinbaozhu@my.swjtu.edu.cn, qifayan@swjtu.edu.cn, xhutang@swjtu.edu.cn).
}
}
\maketitle

\begin{abstract}
 This paper investigates the problem of Secure Multi-party Batch Matrix Multiplication (SMBMM), where a user aims to compute the pairwise products $\mathbf{A}\divideontimes\mathbf{B}\triangleq(\mathbf{A}^{(1)}\mathbf{B}^{(1)},\ldots,\mathbf{A}^{(M)}\mathbf{B}^{(M)})$ of two batch of massive matrices $\mathbf{A}$ and $\mathbf{B}$ that are generated from two sources,  through $N$ honest but curious servers which share some common randomness.
 The matrices $\mathbf{A}$ (resp. $\mathbf{B}$) must be kept secure from any subset of up to $X_{\mathbf{A}}$ (resp. $X_\mathbf{B}$) servers even if they collude, and  the user must not obtain any information about $(\mathbf{A},\mathbf{B})$ beyond the products $\mathbf{A}\divideontimes\mathbf{B}$.
A novel computation strategy for single secure matrix multiplication problem (i.e., the case $M=1$) is first proposed, and then is generalized to the strategy for SMBMM by means  of cross subspace alignment.
The SMBMM strategy focuses on the tradeoff between recovery threshold (the number of successful computing servers that the user needs to wait for), system cost (upload cost, the amount of common randomness, and download cost) and system complexity (encoding, computing, and decoding complexities).
Notably,  compared with the known result by Chen \emph{et al.}, the strategy for the degraded case  $X= X_{\mathbf{A}}=X_{\mathbf{B}}$ achieves
better recovery threshold, amount of common randomness, download cost and decoding complexity when $X$ is less than some parameter threshold, while the performance with respect to other measures remain identical. 

\end{abstract}

\begin{IEEEkeywords}
Distributed matrix multiplication, multi-party computation, security, cross subspace alignment.
\end{IEEEkeywords}

\section{Introduction}\label{Introduction}
Matrix multiplication is one of fundamental building blocks in various engineering applications, for example, big data analysis, machine learning and so on. Massive computation and storage power are usually required in many such applications.
As an efficient solution to process massive datasets,
distributed computing frameworks like MapReduce \cite{MapReduce} and Apache Spark \cite{Spark} enable processing of data sizes at the order of tens of terabytes, by partitioning the large computing task into smaller subtasks and outsourcing them to a set of distributed server nodes, so that they can efficiently circumvent computation and storage barriers of large-scale matrix multiplication.


When scaling out the computation across many distributed servers, the main performance bottleneck of distributed computing is the latency in waiting for slower servers to finish their computing tasks, which are referred to as \textit{stragglers}. It has been demonstrated \cite{Tail1,Tail3} that straggler nodes can be $5$ to $8$ times slower than average and thus cause significant delays in computations. The current approaches to alleviate the impact of stragglers involve injecting some form of computation redundancy across many distributed servers. It was reported that repetition \cite{repetition} and coding techniques \cite{Dutta,Lee,Lee2} 
 are capable of creating and exploiting the computation redundancy to resist the straggling impact. 

In Distributed Matrix Multiplication (DMM) problem where a user wishes to compute the product for two data matrices $\mathbf{A}$ and $\mathbf{B}$ via some distributed servers, the goal is to design an efficient computation strategy with low \emph{recovery threshold}, the number of successful (non-delayed) computing servers that the user needs to wait for, 
 hence mitigating the impact of stragglers. The DMM problem was earlier considered in \cite{Lee,Lee2} by using Maximum Distance Separable (MDS) codes to inject redundancy in data matrices, that is, assigning extra computations to the distributed servers. 
In a recent influential work \cite{Polynomial code}, polynomial codes were used for further improving the recovery threshold by leveraging the algebraic structure of polynomial functions.
Specifically, to multiply the two matrices $\mathbf{A}$ and $\mathbf{B}$, the polynomial codes in \cite{Polynomial code} partition $\mathbf{A}$ row-wise and $\mathbf{B}$ column-wise (row-by-column partitioning), and
generate coded computing redundancy by constructing a polynomial with the partitioning blocks as the coefficients.
Later in \cite{MatDot code}, MatDot codes were introduced to achieve the optimal recovery threshold for column-wise partitioning of matrix $\mathbf{A}$ and row-wise partitioning of $\mathbf{B}$ at the expense of a higher \emph{download cost}, the amount of information to be downloaded from successful servers.
The two concurrent works, both Entangled Polynomial (EP) codes \cite{EP code} and Generalized PolyDot (GPD) codes \cite{GpolyDot}, bridged the gaps of between Polynomial codes \cite{Polynomial code} and MatDot codes \cite{MatDot code} and studied the more general tradeoff between recovery threshold and download cost by arbitrarily partitioning the two data matrices into $m$-by-$p$ and $p$-by-$n$ blocks of equal-size sub-matrices respectively.
{Instead of multiplying two matrices, Jia and Jafar \cite{Jia and Jafar} used arbitrary matrix} partitioning manner to investigate the scenario where the user wishes to compute the pairwise products of two batches of massive matrices $\mathbf{A}=(\mathbf{A}^{(1)},\ldots,\mathbf{A}^{(M)})$ and $\mathbf{B}=(\mathbf{B}^{(1)},\ldots,\mathbf{B}^{(M)})$, also referred to as Batch Matrix Multiplication (BMM). They presented Generalized Cross Subspace Alignment (G-CSA) codes based on the idea of cross subspace alignment \cite{CSA_PIR}. It was shown  that the G-CSA codes for BMM unify and generalize the state-of-art codes for 
DMM problem such as Polynomial codes \cite{Polynomial code}, MatDot codes \cite{MatDot code}, EP codes \cite{EP code} and GPD codes \cite{GpolyDot}. 


On the other hand, coded distributed computing
arises security concerns about outsourcing data information across the distributed servers. It is very desirable to design strategies for  secure DMM that exploit the greater computing power  of  untrustworthy servers while preventing them from learning anything about the data matrices in an information theoretically secure manner.  
The problem of secure DMM was first launched by Tandon \emph{et al.} in \cite{Tandon secure code} for Single Secure Matrix Multiplication (SSMM), which is shown to achieve the optimal download cost for one-sided SSMM (where only one of the data matrices is kept secure).
Subsequently, many works \cite{GSPolyDot code,Rouayheb secure code,Kakar secure code,Kakar and Khristoforov,Gunduz20,EP SMC,Limited-sharing22,Yang secure code} focus on two-sided SSMM (both matrices are kept secure). 
In general, all these works employ the idea of polynomial codes \cite{Polynomial code,EP code,GpolyDot} to construct SSMM strategies based on Shamir's secret sharing scheme \cite{Shamir} to ensure security.
Moreover, distinct partitioning manners of the data matrices are adopted to construct encoded polynomials with the partitioning sub-matrix blocks and the same dimensions of random sub-matrix blocks as the coefficients, while the exponents of the polynomials are carefully chosen to facilitate the interference alignment opportunities. 
Recently, references \cite{Qian Yu} and \cite{batch matrix} combine straggler mitigation and secure computation in BMM setup based on bilinear complexity \cite{Smirnov}, \cite{Strassen} for the multiplication of two matrices.

The problem of Secure Multi-Party Computation (SMPC), first introduced by Yao in \cite{Yao}, focuses on jointly computing an arbitrary polynomial function of some private datasets distributed at the users (parties) under the constraint that each user must not learn any additional information about the datasets beyond the function.
For the secure DMM problem, references \cite{EP SMC,Limited-sharing22,Gunduz20} expanded it under the framework of SMPC,  which allows performing arbitrary polynomial functions on private massive matrices.

The work that is most related to ours is \cite{Chen and Jafar}, which considers a system of Secure Multi-party Batch Matrix Multiplication (SMBMM) including two source nodes, $N$ server nodes and one user node.
There is a link between each source and each server. All of the servers are connected to the user. 
The two sources have access to a batch of $M$ confidential matrices $\mathbf{A}=(\mathbf{A}^{(1)},\ldots,\mathbf{A}^{(M)})$ and $\mathbf{B}=(\mathbf{B}^{(1)},\ldots,\mathbf{B}^{(M)})$, respectively.
The user wishes to efficiently compute the batch of $M$ pairwise products $\mathbf{A}\divideontimes\mathbf{B}=(\mathbf{A}^{(1)}\mathbf{B}^{(1)},\ldots,\mathbf{A}^{(M)}\mathbf{B}^{(M)})$ with the assistance of $N$ distributed computing servers, while ensuring the data matrices secure from any group of up to $X$ colluding server and the user beyond the operation results.

In this paper, we consider the problem of SMBMM in a more general setup where the two data matrices $\mathbf{A}$ and $\mathbf{B}$ have any flexible security levels $X_{\mathbf{A}}$ and $X_{\mathbf{B}}$, i.e., the batch of matrices $\mathbf{A}$ (resp. $\mathbf{B}$) must be kept perfectly secure from all information available to a set of up to $X_{\mathbf{A}}$ (resp. $X_{\mathbf{B}}$) colluding servers, which degrades to the SMBMM problem in \cite{Chen and Jafar} by setting $X=X_{\mathbf{A}}=X_{\mathbf{B}}$.
As an application example for this setup, assume the user wants to train a recommender system based on collaborative filtering \cite{recommender}. In this case,  recommendations are based on the weighted average of the pairwise products of two batches of private matrices.  The matrices represent the training datasets provided by the two sources, which can be the salary information or the medical records, and hence each source needs to keep its datasets secure with self-satisfied security level. 

We will first focus on the SSMM problem (i.e., the case $M=1$), which has received significant recent interest \cite{GSPolyDot code,Rouayheb secure code,Kakar secure code,Kakar and Khristoforov,Gunduz20,EP SMC,Limited-sharing22,Yang secure code}. Similar to these works, in this case the data security requirement against the user is not considered. We then extend our SSMM strategy  to the SMBMM problem.
The main contributions of this paper are two folds:
\begin{enumerate}
  \item For arbitrary matrix partitioning manner, a novel computation strategy for SSMM is proposed to focus on the tradeoff with respect to recovery threshold, system cost for uploads and downloads,  and system complexity for encoding, server computing, and decoding. The strategy achieves a recovery threshold $K=\min\{(m+1)(np+X_{\mathbf{B}})+X_{\mathbf{A}}-X_{\mathbf{B}}-1,(n+1)(mp+X_{\mathbf{A}})+X_{\mathbf{B}}-X_{\mathbf{A}}-1\}$ for any $m,p,n,X_{\mathbf{A}},X_{\mathbf{B}}$.  
  \item The proposed SSMM strategy is extended to SMBMM problem by using the idea of cross subspace alignment, yielding a tradeoff between recovery threshold, system cost (for uploads, the amount of common randomness among servers, and downloads), and system complexity (for encoding, server computing, and decoding).  Accordingly, the recovery threshold $\min\{K',K'' \}$ can be achieved, where $K'=(LG+L-1)mpn+np+X_{\mathbf{A}}+(G+1)(m-1)X_{\mathbf{B}}-1$, $K''=(LG+L-1)mpn+mp+X_{\mathbf{B}}+(G+1)(n-1)X_{\mathbf{A}}-1$ and $M=GL$ for any positive integers $G,L$ with $L>1$.

\end{enumerate}

The rest of this paper is organized as follows. Section \ref{problem statement} introduces the system model. Section \ref{CS_SDMM} presents an SSMM strategy. Section \ref{section:SMBMM} generalizes the SSMM strategy to SMBMM problem. Section \ref{section:comparison} compares the proposed strategies with known results. Finally, the paper is concluded in Section \ref{conclusion}.

\subsubsection*{Notation}
Let boldface capital letters represent matrices.
Denote by $\mathbb{Z}^{+}$ the set of positive integers.
For any $n\in\mathbb{Z}^{+}$ and integer $m$ such that $m<n$, $[n]$ and $[m:n]$ denote the set $\{1,2,\ldots,n\}$ and $\{m,m+1,\ldots,n\}$, respectively.
Given two batches of matrices $\mathbf{A}=(\mathbf{A}^{(1)},\ldots,\mathbf{A}^{(n)})$ and $\mathbf{B}=(\mathbf{B}^{(1)},\ldots,\mathbf{B}^{(n)})$, define $\mathbf{A}\divideontimes\mathbf{B}$ as their pairwise products, i.e., $\mathbf{A}\divideontimes\mathbf{B}=(\mathbf{A}^{(1)}\mathbf{B}^{(1)},\ldots,\mathbf{A}^{(n)}\mathbf{B}^{(n)})$.
Denote by $[\mathbf{0}]_{m\times n}$ the matrix of size $m\times n$ with all the entries being zeros. For a finite set $\mathcal{S}$, $|\mathcal{S}|$ denotes its cardinality.
Define $A_{\Gamma}$ as $\{A_{\gamma_1},\ldots,A_{\gamma_{m}}\}$ for any index set $\Gamma=\{\gamma_1,\ldots,\gamma_{m}\}\subseteq[n]$.
The notation $\mathcal{O}(\cdot)$ denotes the order of the number of arithmetic operations required to perform an algorithm over a finite field, and the order $\widetilde{\mathcal{O}}(m(\log n)^2)$ may be replaced with $\mathcal{O}(m(\log n)^2)$ if the field supports the Fast Fourier Transform (FFT) and with $\mathcal{O}(m(\log n)^2\log\log n)$ otherwise.
The notation $\otimes$ denotes the Kronecker product of two matrices. Let $\mathbf{I}_{n}$ denote $n\times n$ identity matrix and $\mathbf{T}(c_0,c_{1},\ldots,c_{n-1})$ denote the $n\times n$ lower triangular Toeplitz matrix, i.e.,
\begin{IEEEeqnarray*}{c}
\mathbf{T}(c_0,c_{1},\ldots,c_{n-1})=
\left[
\begin{array}{cccc}
c_{0}\\
c_{1}& c_{0} \\
\vdots & \ddots & \ddots  \\
c_{n-1} & \ldots & c_1 & c_{0} \\
\end{array}
\right].
\end{IEEEeqnarray*}


\section{Problem Statement}\label{problem statement}
\begin{figure}[htbp]
\centering
\includegraphics[width=0.55\columnwidth]{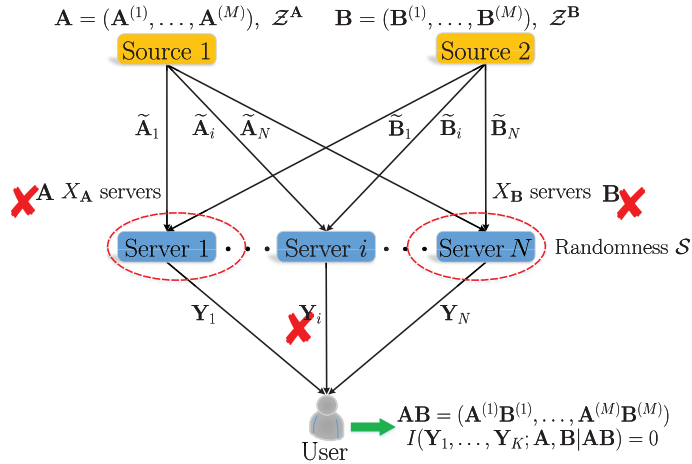}
\caption{\small{System model of secure multi-party batch matrix multiplication.}
}
\label{Fig.1}
\end{figure}

Consider a multi-party computation system including two source nodes, denoted as Source $1$ and Source $2$, $N$ server nodes and one user node. There is no communication link between the two sources but each of them is connected to all the servers. In addition, all the servers are connected to the user, as illustrated in Fig. \ref{Fig.1}. Assume that  all the connected links are error-free and secure. We also assume that the servers are honest but curious, which means that each server follows the protocol and correctly reports any calculations, yet may be curious about the input data and potentially collude to gain information about it.

Source $1$ and Source $2$ respectively have access to $M$ instances of $\lambda\times\xi$ matrices $\mathbf{A}=(\mathbf{A}^{(1)},\ldots,\mathbf{A}^{(M)})$ and $M$ instances of $\xi\times\vartheta$ matrices $\mathbf{B}=(\mathbf{B}^{(1)},\ldots,\mathbf{B}^{(M)})$, for some parameters $M,\lambda,\xi,\vartheta\in\mathbb{Z}^{+}$. All the entries of the matrices are independently and uniformly distributed on $\mathbb{F}_q$ for some given prime power $q$, i.e., in $q$-ary units,
\begin{IEEEeqnarray}{rCl}
H(\mathbf{A}^{(l)})&=&\lambda\xi,\quad H(\mathbf{B}^{(l)})=\xi\vartheta,\quad\forall\,l\in[M],\notag\\
H( \mathbf{A},\mathbf{B})&=&\sum_{l=1}^{M}H(\mathbf{A}^{(l)})+\sum_{l=1}^{M}H(\mathbf{B}^{(l)}). \notag
\end{IEEEeqnarray}


The user is interested in computing  the batch of $M$ pairwise products $\mathbf{A}\divideontimes\mathbf{B}=(\mathbf{A}^{(1)}\mathbf{B}^{(1)},\ldots,$\\$\mathbf{A}^{(M)}\mathbf{B}^{(M)})$ securely using the $N$ distributed servers.
For this purpose, Source $1$ and Source $2$ share some encoding data of matrices $\mathbf{A}$ and $\mathbf{B}$ with each server, respectively.
Since the servers are curious about the data matrices, we impose the constrains of security levels $X_{\mathbf{A}}$ and $X_{\mathbf{B}}$ for $\mathbf{A}$ and $\mathbf{B}$, respectively. That is, any information about matrices $\mathbf{A}$ (resp. $\mathbf{B}$) remains completely unknown to any $X_{\mathbf{A}}$ (resp. $X_{\mathbf{B}}$) of the $N$ servers even if they collude.

Then, each server computes a response according to its available information. Due to the limits of computation and communication resources, some stragglers may fail to respond as required. 
Since the user has no priori knowledge of the identity of the straggler servers, in order to mitigate their influence, the user only downloads the responses from any $K$ servers, i.e., the user can tolerate any subset of up to $N-K$ straggler servers. Consequently, the desired products $\mathbf{A\divideontimes B}$ can be recovered from the responses of any $K$ of the $N$ servers, where $K$ is referred to as \emph{recovery threshold} of the computation strategy. In addition, the user must not gain any information about the data matrices other than the products $\mathbf{A\divideontimes B}$.



Formally, a computation strategy for Secure Multi-party Batch Matrix Multiplication (SMBMM) operates in three phases:
\begin{itemize}
  \item \textbf{Sharing:} To ensure the security of data matrices against the servers, Source $1$ and Source $2$ respectively generate a set of random matrices independently and privately, denoted as $\mathcal{Z}^{\mathbf{A}}$ and $\mathcal{Z}^{\mathbf{B}}$. They are used to encode the matrices $\mathbf{A}$ and $\mathbf{B}$ using  some functions $\bm{f}=(f_1,\ldots,f_{N})$ and $\bm{g}=(g_1,\ldots,g_{N})$ respectively,
        where $f_i$ and $g_{i}$ correspond to server $i$ for any $i\in[N]$. Denote the encoded matrices for server $i$ by $\widetilde{\mathbf{A}}_i$ and $\widetilde{\mathbf{B}}_i$, i.e.,
      \begin{IEEEeqnarray}{c}
       \widetilde{\mathbf{A}}_i=f_{i}(\mathbf{A},\mathcal{Z}^{\mathbf{A}}),\quad\widetilde{\mathbf{B}}_i=g_{i}(\mathbf{B},\mathcal{Z}^{\mathbf{B}}),\quad\forall\,i\in[N], \notag
       \end{IEEEeqnarray}
       which are shared with server $i$. Particularly, the matrices $\mathbf{A}$ and $\mathbf{B}$ must be kept perfectly secure from all information available to a set of up to $X_{\mathbf{A}}$ and $X_{\mathbf{B}}$ colluding servers, respectively, i.e.,
       \begin{IEEEeqnarray}{rCll}
        I(\widetilde{\mathbf{A}}_{\mathcal{X}_{\mathbf{A}}};\mathbf{A})&\overset{(a)}{=}&I(\widetilde{\mathbf{A}}_{\mathcal{X}_{\mathbf{A}}},\widetilde{\mathbf{B}}_{\mathcal{X}_{\mathbf{A}}};\mathbf{A})=0,& \quad \forall\,\mathcal{X}_{\mathbf{A}}\subseteq[N],|\mathcal{X}_{\mathbf{A}}|=X_{\mathbf{A}}, \label{sec const}\\
        I(\widetilde{\mathbf{B}}_{\mathcal{X}_{\mathbf{B}}};\mathbf{B})&\overset{(b)}{=}&I(\widetilde{\mathbf{A}}_{\mathcal{X}_{\mathbf{B}}},\widetilde{\mathbf{B}}_{\mathcal{X}_{\mathbf{B}}};\mathbf{B})=0, &\quad \forall\,\mathcal{X}_{\mathbf{B}}\subseteq[N],|\mathcal{X}_{\mathbf{B}}|=X_{\mathbf{B}}, \label{sec const2}
       \end{IEEEeqnarray}
        where $(a)$ and $(b)$ are due to the fact that the information available to the two sources are independent and thus $I(\mathbf{A}_{[N]},\widetilde{\mathbf{A}}_{[N]};\mathbf{B}_{[N]},\widetilde{\mathbf{B}}_{[N]})=0$.
  \item \textbf{Computation:} In order to ensure the security of data matrices against the user beyond the desired products $\mathbf{A\divideontimes B}$, we assume that the servers share a set of common random
variables $\mathcal{S}$ that are generated independently of the data matrices $\mathbf{A},\mathbf{B}$ and are unknown to the user.
This is also the only cost for imposing the data security constrain against the user.
Then, server $i$ prepares a response $\mathbf{Y}_{i}$ for the user, which is a deterministic function of all the information available to the server, i.e.,
      \begin{IEEEeqnarray}{c}
       H(\mathbf{Y}_{i}|\widetilde{\mathbf{A}}_i,\widetilde{\mathbf{B}}_i,\mathcal{S})=0,\quad\forall\,i\in[N]. \notag
       \end{IEEEeqnarray}
  \item \textbf{Reconstruction:} 
      The final products should be completely disclosed by the collection of responses of any $K$ servers, i.e.,
\begin{IEEEeqnarray}{c}
       H(\mathbf{A\divideontimes B}|\mathbf{Y}_{\mathcal{K}})=0,\quad\forall\,\mathcal{K}\subseteq[N],|\mathcal{K}|=K, \notag
       \end{IEEEeqnarray}
  while secure multi-party computation imposes the additional constraint keeping the data matrices secure from the user beyond the interested products, i.e.,
  \begin{IEEEeqnarray}{c}\label{security for user}
  I(\mathbf{Y}_{\mathcal{K}};\mathbf{A},\mathbf{B}|\mathbf{A\divideontimes B})=0, \quad\forall\,\mathcal{K}\subseteq[N],|\mathcal{K}|=K.
  \end{IEEEeqnarray}
\end{itemize}

The performance of a computation strategy for SMBMM are measured by three quantities:
\begin{enumerate}
  \item[1.] The recovery threshold $K$, which is the number of server  responses  that the user needs to collect in order to recover the desired products $\mathbf{A\divideontimes B}$,  is preferred to be as small as possible.
  \item[2.] The system cost, which is comprised of three parts: 
  the upload cost for the data matrices $\mathbf{A}$ and $\mathbf{B}$, defined as
    \begin{IEEEeqnarray}{c}
    U_{\mathbf{A}}\triangleq\frac{\sum_{i=1}^{N}H(\widetilde{\mathbf{A}}_i)}{M\lambda\xi}, \quad 
    U_{\mathbf{B}}\triangleq\frac{\sum_{i=1}^{N}H(\widetilde{\mathbf{B}}_i)}{M\xi\vartheta},\label{upload cost:B}
    \end{IEEEeqnarray}
  the amount of common randomness and download cost, defined as
    \begin{IEEEeqnarray}{c}
    \rho\triangleq
    \frac{H(\mathcal{S})}{M\lambda\vartheta}, \quad
    D\triangleq
    \max\limits_{\mathcal{K}:\mathcal{K}\subseteq[N],|\mathcal{K}|=K}\frac{\sum_{i\in\mathcal{K}}H({\mathbf{Y}}_i)}{M\lambda\vartheta}.\label{download cost}
    \end{IEEEeqnarray}
  \item[3.] The system complexity, which includes the complexities of encoding, server computation and decoding. The encoding complexities $\mathcal{C}_{\mathbf{A}},\mathcal{C}_{\mathbf{B}}$ at Source $1$ and Source $2$ are defined as the \emph{order} of the number of arithmetic operations required to compute the encoding functions $\bm{f}$ and $\bm{g}$, respectively, each normalized by $M$. The server computation complexity $\mathcal{C}_{s}$ is defined as  the \emph{order} of the number of arithmetic operations required to generate the response $\mathbf{Y}_i$, maximized over $i\in[N]$ and normalized by $M$.
         And the decoding complexity $\mathcal{C}_d$ at the user is defined as the \emph{order} of the number of arithmetic operations required to decode the desired products $\mathbf{A\divideontimes B}$ from the answers of responsive servers $\mathcal{K}$, maximized over $\mathcal{K}\subseteq[N],|\mathcal{K}|=K$ and normalized by $M$.

\end{enumerate}

The objective of this paper is to design computation strategies with flexible tradeoff between recovery threshold, system cost and system complexity for SMBMM. 
\begin{Remark}
In the SMBMM problem, we allow the security levels $X_{\mathbf{A}}$ and $X_{\mathbf{B}}$ to be arbitrary size, which contains the work considered by Chen \emph{et al.} \cite{Chen and Jafar} as a special case by setting $X_{\mathbf{A}}=X_{\mathbf{B}}$.
\end{Remark}

\section{Computation Strategies for Single Secure Matrix Multiplication}\label{CS_SDMM}
In this section, we consider the traditional problem of SSMM, and accordingly, a novel computation strategy is proposed, which will be generalized to the strategy for SMBMM in the subsequent sections. As mentioned in Section \ref{Introduction}, the security constraint \eqref{security for user} against the user is not imposed and hence the common randomness $\mathcal{S}$ is not necessary, i.e., its amount $\rho$ can be set to be $0$. In this case, the user wishes to compute the single product $\mathbf{C}=\mathbf{AB}$ of matrix $\mathbf{A}$ with security level $X_{\mathbf{A}}$ and matrix $\mathbf{B}$ with security level $X_{\mathbf{B}}$.




We first introduce a lemma that is used for keeping  the data matrices secure from the servers.
\begin{Lemma}[Secret Sharing \cite{Shamir}]\label{security proof}
For any parameters $L,X,\lambda,\vartheta\in\mathbb{Z}^{+}$ and a prime power $q$ with $q>N$, let $\mathbf{W}_{1},\ldots,\mathbf{W}_{L}\in\mathbb{F}_q^{\lambda\times\vartheta}$ be $L$ secrets, and $\mathbf{Z}_{1},\ldots,\mathbf{Z}_{X}$ be $X$ random matrices of the same dimensions as secrets whose entries are chosen independently and uniformly from $\mathbb{F}_q$. Let $\alpha_1,\ldots,\alpha_N$ be $N$ distinct numbers from $\mathbb{F}_q$.
Define a function of $\alpha$ as
\begin{IEEEeqnarray}{c}
\mathbf{Y}(\alpha)=\mathbf{W}_{1}u_1(\alpha)+\ldots+\mathbf{W}_{L}u_L(\alpha)+\mathbf{Z}_{1}v_1(\alpha)+\ldots+\mathbf{Z}_{X}v_X(\alpha),\notag
\end{IEEEeqnarray}
where $u_1(\alpha),\ldots,u_L(\alpha),v_1(\alpha),\ldots,v_X(\alpha)$ are the deterministic function of $\alpha$.
If the matrix
\begin{IEEEeqnarray}{c}\label{constructed matrix}
\mathbf{V}=
\left[
  \begin{array}{ccc}
    v_1(\alpha_{i_1})  & \ldots & v_X(\alpha_{i_1}) \\
    \vdots & \ddots & \vdots \\
    v_1(\alpha_{i_X})  & \ldots & v_X(\alpha_{i_X}) \\
  \end{array}
\right]_{X\times X} \notag
\end{IEEEeqnarray}
is non-singular over $\mathbb{F}_q$ for any $\mathcal{X}=\{i_1,\ldots,i_X\}\subseteq[N]$ with $|\mathcal{X}|=X$, then the $X$ values $\mathbf{Y}(\alpha_{i_1}),\ldots,\mathbf{Y}(\alpha_{i_X})$  can not learn any information about matrices $\mathbf{W}_{1},\ldots,\mathbf{W}_{L}$, i.e.,
\begin{IEEEeqnarray}{c}\label{security formulate}
I(\mathbf{Y}(\alpha_{i_1}),\ldots,\mathbf{Y}(\alpha_{i_X});\mathbf{W}_{1},\ldots,\mathbf{W}_{L})=0,\quad\forall\,\mathcal{X}=\{i_1,\ldots,i_X\}\subseteq[N],|\mathcal{X}|=X.\notag
\end{IEEEeqnarray}
\end{Lemma}


Let $m,p,n\in\mathbb{Z}^{+}$ be the partitioning parameters of data matrices. Then, $\mathbf{A}$ and $\mathbf{B}$ are partitioned into $m\times p$ and $p\times n$ equal-size sub-matrices, respectively, as shown below.
\begin{IEEEeqnarray}{c}\label{partition3214}
\mathbf{A}=
\left[
  \begin{array}{ccc}
    \mathbf{A}_{1,1}  & \ldots & \mathbf{A}_{1,p} \\
    \vdots & \ddots & \vdots \\
    \mathbf{A}_{m,1}  & \ldots & \mathbf{A}_{m,p} \\
  \end{array}
\right], \quad
\mathbf{B}=
\left[
  \begin{array}{ccc}
    \mathbf{B}_{1,1} & \ldots & \mathbf{B}_{1,n} \\
    \vdots  & \ddots & \vdots \\
    \mathbf{B}_{p,1}  & \ldots & \mathbf{B}_{p,n} \\
  \end{array}
\right],
\end{IEEEeqnarray}
where $\mathbf{A}_{k,l}\in\mathbb{F}_q^{\frac{\lambda}{m}\times\frac{\xi}{p}}$ for any $k\in[m],l\in[p]$ and $\mathbf{B}_{l,j}\in\mathbb{F}_q^{\frac{\xi}{p}\times\frac{\vartheta}{n}}$ for any $l\in[p],j\in[n]$.


Accordingly, the desired matrix product $\mathbf{C}=\mathbf{AB}$ involves a total of $mn$ linear combinations of products of sub-matrices, i.e.,
\begin{IEEEeqnarray}{c}\label{EP code Result}
\mathbf{C}=\mathbf{AB}=
\left[
  \begin{array}{ccc}
    \mathbf{C}_{1,1} & \ldots & \mathbf{C}_{1,n} \\
    \vdots & \ddots & \vdots \\
    \mathbf{C}_{m,1}  & \ldots & \mathbf{C}_{m,n} \\
  \end{array}
\right],
\end{IEEEeqnarray}
where $\mathbf{C}_{k,j}=\sum_{l=1}^{p}\mathbf{A}_{k,l}\mathbf{B}_{l,j}$ for $k\in[m],j\in[n]$.

\subsection{An Illustrative Example for SSMM}\label{example:1}
We illustrate how the SSMM strategy works through a detailed example for the parameters $m=2,p=3,n=2,X_{\mathbf{A}}=2,X_{\mathbf{B}}=3$.
In the following subsection, we will describe the general strategy, but this example suffices to convey the essential ingredients behind the strategy.
The matrices $\mathbf{A},\mathbf{B}$ are partitioned as follows.
\begin{IEEEeqnarray}{c}\label{desired product:exam222}
\mathbf{A}=
\left[
  \begin{array}{ccc}
    \mathbf{A}_{1,1} & \mathbf{A}_{1,2} & \mathbf{A}_{1,3} \\
    \mathbf{A}_{2,1} & \mathbf{A}_{2,2} & \mathbf{A}_{2,3} \\
  \end{array}
\right], \quad
\mathbf{B}=
\left[
  \begin{array}{cc}
    \mathbf{B}_{1,1} & \mathbf{B}_{1,2} \\
    \mathbf{B}_{2,1} & \mathbf{B}_{2,2} \\
    \mathbf{B}_{3,1} & \mathbf{B}_{3,2} \\
  \end{array}
\right],
\end{IEEEeqnarray}
where $\mathbf{A}_{k,l}\in\mathbb{F}_q^{\frac{\lambda}{2}\times\frac{\xi}{3}}$ for any $k\in[2],l\in[3]$ and $\mathbf{B}_{l,j}\in\mathbb{F}_q^{\frac{\xi}{3}\times\frac{\vartheta}{2}}$ for any $l\in[3],j\in[2]$.
The objective of the user is to calculate
\begin{IEEEeqnarray}{c}\label{desired product:exam}
\mathbf{C}=\mathbf{AB}=
\left[
  \begin{array}{cc}
    \mathbf{C}_{1,1} & \mathbf{C}_{1,2} \\
    \mathbf{C}_{2,1} & \mathbf{C}_{2,2} \\
  \end{array}
\right],
\end{IEEEeqnarray}
where $\mathbf{C}_{k,j}=\mathbf{A}_{k,1}\mathbf{B}_{1,j}+\mathbf{A}_{k,2}\mathbf{B}_{2,j}+\mathbf{A}_{k,3}\mathbf{B}_{3,j}$ for any $k\in[2],j\in[2]$.


{To this end, the two sources first employ MatDot codes \cite{MatDot code} to encode the partitioning sub-matrices in row $k$ of $\mathbf{A}$ and column $j$ of $\mathbf{B}$ as follows.
\begin{IEEEeqnarray}{rCl}\label{matdot codes:2}
\mathbf{A}_{k}=\mathbf{A}_{k,1}+\mathbf{A}_{k,2}\alpha+\mathbf{A}_{k,3}\alpha^2,\quad \mathbf{B}_{j}=\mathbf{B}_{1,j}\alpha^2+\mathbf{B}_{2,j}\alpha+\mathbf{B}_{3,j}.\notag
\end{IEEEeqnarray}
Then, we have
\begin{IEEEeqnarray}{rCl}\label{example:product}
\mathbf{H}_{k,j}=\mathbf{A}_{k}\mathbf{B}_{j}=\mathbf{H}_{0}^{k,j}+\mathbf{H}_{1}^{k,j}\alpha+\mathbf{H}_{2}^{k,j}\alpha^2+\mathbf{H}_{3}^{k,j}\alpha^3+\mathbf{H}_{4}^{k,j}\alpha^4,
\end{IEEEeqnarray}
where $\mathbf{H}_{r}^{k,j}$ is the coefficient of $\alpha^{r}$ for $r\in[0:4]$. It is easy to verify $\mathbf{H}_{2}^{k,j}=\mathbf{C}_{k,j}$.

To keep the data matrices secure from the servers, the following polynomials are employed to further encode $\mathbf{A}_{k}$ and $\mathbf{B}_{j}$:
\begin{IEEEeqnarray}{rCl}
\widetilde{\mathbf{A}}(\alpha)&=&\mathbf{A}_{1}\alpha^{s_1}+\mathbf{A}_{2}\alpha^{s_2}+\mathbf{Z}_1^{\mathbf{A}}\alpha^{s_3}+\mathbf{Z}_2^{\mathbf{A}}\alpha^{s_3+1},\notag\\
\widetilde{\mathbf{B}}(\alpha)&=&\mathbf{B}_{1}\alpha^{t_1}+\mathbf{B}_{2}\alpha^{t_2}+\mathbf{Z}_1^{\mathbf{B}}\alpha^{t_3}+\mathbf{Z}_2^{\mathbf{B}}\alpha^{t_3+1}+\mathbf{Z}_3^{\mathbf{B}}\alpha^{t_3+2},\notag
\end{IEEEeqnarray}
where the exponents of the $\alpha$ will be determined shortly, and $\mathbf{Z}_1^{\mathbf{A}},\mathbf{Z}_2^{\mathbf{A}}\in\mathbb{F}_q^{\frac{\lambda}{2}\times\frac{\xi}{3}}$ and $\mathbf{Z}_1^{\mathbf{B}},\mathbf{Z}_2^{\mathbf{B}},\mathbf{Z}_3^{\mathbf{B}}\in\mathbb{F}_q^{\frac{\xi}{3}\times\frac{\vartheta}{2}}$ are the matrices picked independently and uniformly at random with entries in $\mathbb{F}_q$.
Moreover, the random noises $\mathbf{Z}_1^{\mathbf{A}},\mathbf{Z}_2^{\mathbf{A}}$ in $\widetilde{\mathbf{A}}(\alpha)$ correspond to consecutive exponents $s_3,s_3+1$ and so does $\widetilde{\mathbf{B}}(\alpha)$, which keep the data matrices $\mathbf{A},\mathbf{B}$ secure from servers.

The user will recover the desired product $\mathbf{C}_{k,j}$ by interpolating the polynomial $\widetilde{\mathbf{A}}(\alpha)\cdot\widetilde{\mathbf{B}}(\alpha)$.
Specifically, let $\alpha_1,\ldots,\alpha_N$ be $N$ distinct non-zero elements in $\mathbb{F}_q$.
For any $i\in[N]$, Source $1$ and Source $2$ respectively share $\widetilde{\mathbf{A}}(\alpha_i)$ and $\widetilde{\mathbf{B}}(\alpha_i)$ with server $i$ where the server in return calculates the product $\widetilde{\mathbf{A}}(\alpha_i)\cdot\widetilde{\mathbf{B}}(\alpha_i)$ and responds the result back to the user.
These evaluations will suffice to interpolate $\widetilde{\mathbf{A}}(\alpha)\cdot\widetilde{\mathbf{B}}(\alpha)$. 

To ensure decodability, we want to choose the exponents of the $\alpha$ such that $\mathbf{C}_{1,1},\mathbf{C}_{1,2},\mathbf{C}_{2,1},\mathbf{C}_{2,2}$ should be separated from each other and also from all the other undesired components, while the undesired components (interference) should be aligned such that the degree of $\widetilde{\mathbf{A}}(\alpha)\cdot\widetilde{\mathbf{B}}(\alpha)$ is minimized and thus minimize the recovery threshold.

Next, we show the interference alignment opportunities by describing the steps choosing a set of appropriate exponents of the $\alpha$, subject to the above constraints.  The product polynomial $\widetilde{\mathbf{A}}(\alpha)\cdot\widetilde{\mathbf{B}}(\alpha)$ is expanded by using multiplicative distribution law as follows.
\begin{IEEEeqnarray}{rCl}
\widetilde{\mathbf{A}}(\alpha)\cdot \widetilde{\mathbf{B}}(\alpha)&=&\underbrace{\mathbf{A}_{1}(\mathbf{B}_{1}\alpha^{t_1}+\mathbf{B}_{2}\alpha^{t_2}+\mathbf{Z}_1^{\mathbf{B}}\alpha^{t_3}+\mathbf{Z}_2^{\mathbf{B}}\alpha^{t_3+1}+\mathbf{Z}_3^{\mathbf{B}}\alpha^{t_3+2})}_{=p_1(\alpha)}\alpha^{s_1}+\label{align rule:1}\notag\\
&&\underbrace{\mathbf{A}_{2}(\mathbf{B}_{1}\alpha^{t_1}+\mathbf{B}_{2}\alpha^{t_2}+\mathbf{Z}_1^{\mathbf{B}}\alpha^{t_3}+\mathbf{Z}_2^{\mathbf{B}}\alpha^{t_3+1}+\mathbf{Z}_3^{\mathbf{B}}\alpha^{t_3+2})}_{=p_2(\alpha)}\alpha^{s_2}+\notag\\
&&\underbrace{\mathbf{Z}_1^{\mathbf{A}}(\mathbf{B}_{1}\alpha^{t_1}+\mathbf{B}_{2}\alpha^{t_2}+\mathbf{Z}_1^{\mathbf{B}}\alpha^{t_3}+\mathbf{Z}_2^{\mathbf{B}}\alpha^{t_3+1}+\mathbf{Z}_3^{\mathbf{B}}\alpha^{t_3+2})}_{=p_3(\alpha)}\alpha^{s_3}+\notag\\
&&\underbrace{\mathbf{Z}_2^{\mathbf{A}}(\mathbf{B}_{1}\alpha^{t_1}+\mathbf{B}_{2}\alpha^{t_2}+\mathbf{Z}_1^{\mathbf{B}}\alpha^{t_3}+\mathbf{Z}_2^{\mathbf{B}}\alpha^{t_3+1}+\mathbf{Z}_3^{\mathbf{B}}\alpha^{t_3+2})}_{=p_4(\alpha)}\alpha^{s_3+1}.\notag
\end{IEEEeqnarray}

In general, to minimize the degree of $\widetilde{\mathbf{A}}(\alpha)\cdot \widetilde{\mathbf{B}}(\alpha)$, we need to align the interference from the undesired components in $\widetilde{\mathbf{A}}(\alpha)\cdot\widetilde{\mathbf{B}}(\alpha)$ such that they occur in the exponents as low as possible.
Thus, we initially set $t_1=0,s_1=0$, which enables the desired coefficient $\mathbf{H}_2^{1,1}=\mathbf{C}_{1,1}$ in $\mathbf{A}_1\mathbf{B}_1$ occur in the exponent $\alpha^2$ by \eqref{example:product}.
To further obtain the desired coefficient $\mathbf{H}_2^{1,2}=\mathbf{C}_{1,2}$ in $\mathbf{A}_1\mathbf{B}_2$ and align the interference from the undesired components in $\mathbf{A}_1\mathbf{B}_1$ and $\mathbf{A}_1\mathbf{B}_2$, $t_2$ is set to be $3$ such that $\mathbf{H}_2^{1,2}$ occurs in the exponent $\alpha^5$, and $\mathbf{H}_{3}^{1,1}$ and $\mathbf{H}_{4}^{1,1}$ in $\mathbf{A}_1\mathbf{B}_1$ are aligned with $\mathbf{H}_{0}^{1,2}$ and $\mathbf{H}_{1}^{1,2}$ in $\mathbf{A}_1\mathbf{B}_2$ on the exponents $\alpha^3$ and $\alpha^4$, respectively.
Since there are no desired components in $\mathbf{A}_1\mathbf{Z}_1^{\mathbf{B}},\mathbf{A}_1\mathbf{Z}_2^{\mathbf{B}},\mathbf{A}_1\mathbf{Z}_3^{\mathbf{B}}$, they only need to be aligned with the undesired components in $\mathbf{A}_1\mathbf{B}_2$. Thus, we set $t_3=6$. One can observe such alignment opportunities by expanding $p_1(\alpha)$:
\begin{IEEEeqnarray}{rCl}
p_1(\alpha)&=&\mathbf{A}_{1}(\mathbf{B}_{1}+\mathbf{B}_{2}\alpha^{3}+\mathbf{Z}_1^{\mathbf{B}}\alpha^{6}+\mathbf{Z}_2^{\mathbf{B}}\alpha^{7}+\mathbf{Z}_3^{\mathbf{B}}\alpha^{8}) \notag\\
&=&\mathbf{H}_{0}^{1,1}+\mathbf{H}_{1}^{1,1}\alpha+\underbrace{\mathbf{H}_{2}^{1,1}}_{\text{Desired}}\alpha^2+(\underbrace{\mathbf{H}_{3}^{1,1}+\mathbf{H}_{0}^{1,2}}_{\text{Alignment}})\alpha^3+(\underbrace{\mathbf{H}_{4}^{1,1}+\mathbf{H}_{1}^{1,2}}_{\text{Alignment}})\alpha^4+\underbrace{\mathbf{H}_{2}^{1,2}}_{\text{Desired}}\alpha^5\label{align rule:11}\\
&&+(\underbrace{\mathbf{H}_{3}^{1,2}+\mathbf{A}_{1,1}\mathbf{Z}_1^{\mathbf{B}}}_{\text{Alignment}})\alpha^6+(\underbrace{\mathbf{H}_{4}^{1,2}+\mathbf{A}_{1,2}\mathbf{Z}_1^{\mathbf{B}}+\mathbf{A}_{1,1}\mathbf{Z}_2^{\mathbf{B}}}_{\text{Alignment}})\alpha^7\\
&&+(\underbrace{\mathbf{A}_{1,3}\mathbf{Z}_1^{\mathbf{B}}+\mathbf{A}_{1,2}\mathbf{Z}_2^{\mathbf{B}}+\mathbf{A}_{1,1}\mathbf{Z}_3^{\mathbf{B}}}_{\text{Alignment}})\alpha^8+(\underbrace{\mathbf{A}_{1,3}\mathbf{Z}_2^{\mathbf{B}}+\mathbf{A}_{1,2}\mathbf{Z}_3^{\mathbf{B}}}_{\text{Alignment}})\alpha^9+\mathbf{A}_{1,3}\mathbf{Z}_3^{\mathbf{B}}\alpha^{10}.\label{align rule:22}
\end{IEEEeqnarray}

Similarly, to avoid overlaps of the desired coefficients $\mathbf{C}_{2,1},\mathbf{C}_{2,2}$ in $p_2(\alpha)$ with other components while simultaneously aligning interference from the remaining components in $p_2(\alpha)$ and all the components in $p_3(\alpha),p_4(\alpha)$, we choose $s_2=9,s_3=15$.
As a result, the encoding functions of matrices $\mathbf{A}$ and $\mathbf{B}$ are constructed as
\begin{IEEEeqnarray}{rCl}
\widetilde{\mathbf{A}}(\alpha)&=& (\mathbf{A}_{1,1}+\mathbf{A}_{1,2}\alpha+\mathbf{A}_{1,3}\alpha^2)+(\mathbf{A}_{2,1}+\mathbf{A}_{2,2}\alpha+\mathbf{A}_{2,3}\alpha^2)\alpha^{9}+\mathbf{Z}_1^{\mathbf{A}}\alpha^{15}+\mathbf{Z}_2^{\mathbf{A}}\alpha^{16}, \label{sharing:1}\\
\widetilde{\mathbf{B}}(\alpha)&=& (\mathbf{B}_{1,1}\alpha^2+\mathbf{B}_{2,1}\alpha+\mathbf{B}_{3,1})+(\mathbf{B}_{1,2}\alpha^2+\mathbf{B}_{2,2}\alpha+\mathbf{B}_{3,2})\alpha^3+\mathbf{Z}_1^{\mathbf{B}}\alpha^{6}+\mathbf{Z}_2^{\mathbf{B}}\alpha^{7}+\mathbf{Z}_3^{\mathbf{B}}\alpha^{8}. \IEEEeqnarraynumspace \label{sharing:2}
\end{IEEEeqnarray}
It is easy to check that the product $\widetilde{\mathbf{A}}(\alpha)\cdot \widetilde{\mathbf{B}}(\alpha)$ contains $\mathbf{C}_{1,1},\mathbf{C}_{1,2},\mathbf{C}_{2,1},\mathbf{C}_{2,2}$ as coefficients by expanding $\widetilde{\mathbf{A}}(\alpha)\cdot \widetilde{\mathbf{B}}(\alpha)=p_1(\alpha)+p_2(\alpha)\alpha^9+p_3(\alpha)\alpha^{15}+p_4(\alpha)\alpha^{16}$. Thus, the recovery threshold is $K=\deg(\widetilde{\mathbf{A}}(\alpha)\cdot \widetilde{\mathbf{B}}(\alpha))+1=25$.
}

\subsection{Computation Strategy for SSMM}\label{construction_SDMM}
In this subsection, we formally present the general construction for the computation strategy of SSMM. In essence, the strategy generalizes and improves the aligned secret sharing scheme (A3S) \cite{Kakar secure code} to provide flexible tradeoff with respect to the performance for arbitrary partitioning manner and security levels of the data matrices.
Both A3S codes and our SSMM strategy are constructed based on Polynomial codes \cite{Polynomial code} and Shamir's secret sharing scheme \cite{Shamir}.
However, when considering arbitrary partitioning and security levels, the exponents in the encoding functions of the data matrices need to be more elaborately chosen to create interference alignment opportunities when recovering the desired product, as shown in \eqref{align rule:11}-\eqref{align rule:22} in the above example.

Recall from \eqref{EP code Result} that to recover $\mathbf{AB}$, the user is interested in the products of sub-matrices $\mathbf{C}_{k,j},k\in[m],j\in[n]$.
In general, the key technique in the strategy is to choose appropriate exponents of the $\alpha$ in the encoding functions $\widetilde{\mathbf{A}}(\alpha)$ and $\widetilde{\mathbf{B}}(\alpha)$  such that the following criteria are satisfied:
\begin{itemize}
\item[C1] The polynomial $\mathbf{H}(\alpha)=\widetilde{\mathbf{A}}(\alpha)\cdot\widetilde{\mathbf{B}}(\alpha)$ contains the desired sub-products $\mathbf{C}_{k,j},k\in[m],j\in[n]$ as coefficients. Meanwhile, all of the exponents corresponding to these desired terms are pairwise distinct and are also distinct from the exponents corresponding to all the other terms appearing in $\mathbf{H}(\alpha)$. This guarantees decodability.
\item[C2] Except for the $mn$ desired terms in $\mathbf{H}(\alpha)$, the rest are interference and thus their corresponding exponents should be overlapped as much as possible. This will allow us to align the interference from the undesired terms  at utmost and thus minimize the recovery threshold.
\item[C3] In $\widetilde{\mathbf{A}}(\alpha)$ and $\widetilde{\mathbf{B}}(\alpha)$, the exponents of the terms with random noise matrices are numerically continuous. This ensures the security of data matrices against the servers by Lemma \ref{security proof} and the non-singular property of Vandermonde matrix.
\end{itemize}



\textbf{Sharing: }In order to keep $\mathbf{A}$ secure from any $X_{\mathbf{A}}$ colluding servers, Source $1$ generates $X_{\mathbf{A}}$ random matrices $\mathbf{Z}_{1}^{\mathbf{A}},\ldots,\mathbf{Z}_{X_{\mathbf{A}}}^{\mathbf{A}}$ distributed independently and uniformly on $\mathbb{F}_q^{\frac{\lambda}{m}\times\frac{\xi}{p}}$. Similarly, Source $2$ generates  $X_{\mathbf{B}}$ random matrices $\mathbf{Z}_{1}^{\mathbf{B}},\ldots,\mathbf{Z}_{X_{\mathbf{B}}}^{\mathbf{B}}$ distributed independently and uniformly on $\mathbb{F}_q^{\frac{\xi}{p}\times\frac{\vartheta}{n}}$.
Based on the  criteria C1-C3, construct the encoding functions of data matrices $\mathbf{A}$ and $\mathbf{B}$ as:
\begin{IEEEeqnarray}{rCl}
\widetilde{\mathbf{A}}(\alpha)&=&\sum\limits_{k=1}^{m}\left(\sum\limits_{l=1}^{p}\mathbf{A}_{k,l}\alpha^{l-1}\right)\alpha^{(k-1)(np+X_{\mathbf{B}})}+\sum\limits_{x=1}^{X_{\mathbf{A}}}\mathbf{Z}_{x}^{\mathbf{A}}\alpha^{{(m-1)(np+X_{\mathbf{B}})+np+(x-1)}}, \label{encoded marix:C1}\\
\widetilde{\mathbf{B}}(\alpha)&=&\sum\limits_{j=1}^{n}\left(\sum\limits_{l=1}^{p}\mathbf{B}_{l,j}\alpha^{p-l}\right)\alpha^{(j-1)p}+\sum\limits_{x=1}^{X_{\mathbf{B}}}\mathbf{Z}_{x}^{\mathbf{B}}\alpha^{np+(x-1)}, \label{encoded marix:C2}
\end{IEEEeqnarray}
Obviously, the encoding functions $\widetilde{\mathbf{A}}(\alpha)$ and $\widetilde{\mathbf{B}}(\alpha)$ satisfy the criterion C3.

Let $\alpha_1,\alpha_2,\ldots,\alpha_N$ be $N$ distinct non-zero elements in $\mathbb{F}_q$.
Then Source $1$ and Source $2$ share the evaluations of $\widetilde{\mathbf{A}}(\alpha)$ and $\widetilde{\mathbf{B}}(\alpha)$ at $\alpha=\alpha_i$ with server $i\in[N]$, respectively. 

\textbf{Computation: }The $i$-th server computes the product $\widetilde{\mathbf{A}}(\alpha_i)\cdot\widetilde{\mathbf{B}}(\alpha_i)$ of the two received encoding sub-matrices, and sends it to the user on successful completion. The degree of $\widetilde{\mathbf{A}}(\alpha)\cdot\widetilde{\mathbf{B}}(\alpha)$ is given by
\begin{IEEEeqnarray}{c}\label{delta:def}
\delta\triangleq(m+1)(np+X_{\mathbf{B}})+X_{\mathbf{A}}-X_{\mathbf{B}}-2.
\end{IEEEeqnarray}
 Thus the polynomial can be written as
\begin{IEEEeqnarray}{c}\label{ply:product}
\mathbf{H}(\alpha)=\sum\limits_{r=0}^{\delta}\mathbf{H}_r\alpha^r=\widetilde{\mathbf{A}}(\alpha)\cdot\widetilde{\mathbf{B}}(\alpha),
\end{IEEEeqnarray}
where $\mathbf{H}_r$ is the coefficient of $\alpha^r$ in the polynomial $\widetilde{\mathbf{A}}(\alpha)\cdot\widetilde{\mathbf{B}}(\alpha)$. That is, the response of server $i$ is equivalent to evaluating $\mathbf{H}(\alpha)$ at $\alpha=\alpha_i$ for any $i\in[N]$.

\textbf{Reconstruction: } It is easy to prove that
\begin{IEEEeqnarray}{c}\label{desired:coeffi}
\mathbf{H}_{(k-1)(np+X_{\mathbf{B}})+jp-1}=\sum\limits_{l=1}^{p}\mathbf{A}_{k,l}\mathbf{B}_{l,j}=\mathbf{C}_{k,j},\quad\forall\,k\in[m],j\in[n].
\end{IEEEeqnarray}
Thus the desired products $\mathbf{C}_{k,j},k\in[m],j\in[n]$ are separated from the remaining undesired components and the criterion C1 is satisfied. Further, one can verify the interference alignment criterion C2 by observing the remaining undesired coefficients.

Thus, if the user is able to collect all the coefficients of $\mathbf{H}(\alpha)$ in \eqref{ply:product}, then the final product $\mathbf{C}=\mathbf{AB}$ can be reconstructed. Since the degree of polynomial $\mathbf{H}(\alpha)$ is $\delta$, the evaluation of $\mathbf{H}(\alpha)$ at any $\delta+1$ distinct points is sufficient to recover all the coefficients of $\mathbf{H}(\alpha)$ by using Lagrange interpolation rules.
Therefore, the user only needs to wait for the responses of any $\delta+1=(m+1)(np+X_{\mathbf{B}})+X_{\mathbf{A}}-X_{\mathbf{B}}-1$ of the $N$ servers because the $N$ evaluation points $\alpha_1,\ldots,\alpha_N$ are distinct, i.e., the computation strategy achieves the recovery threshold $K=(m+1)(np+X_{\mathbf{B}})+X_{\mathbf{A}}-X_{\mathbf{B}}-1$.

Intriguingly, if we exchange the two encoding functions of $\mathbf{A}$ and $\mathbf{B}$ in \eqref{encoded marix:C1} and \eqref{encoded marix:C2}, i.e., let the functions of $\mathbf{A}$ and $\mathbf{B}$ be
\begin{IEEEeqnarray}{rCl}
\widetilde{\mathbf{A}}(\alpha)&=&\sum\limits_{k=1}^{m}\left(\sum\limits_{l=1}^{p}\mathbf{A}_{k,l}\alpha^{l-1}\right)\alpha^{(k-1)p}+\sum\limits_{x=1}^{X_{\mathbf{A}}}\mathbf{Z}_{x}^{\mathbf{A}}\alpha^{mp+(x-1)},\label{encoding:function:1}\\
\widetilde{\mathbf{B}}(\alpha)&=&\sum\limits_{j=1}^{n}\left(\sum\limits_{l=1}^{p}\mathbf{B}_{l,j}\alpha^{p-l}\right)\alpha^{(j-1)(mp+X_{\mathbf{A}})}+\sum\limits_{x=1}^{X_{\mathbf{B}}}\mathbf{Z}_{x}^{\mathbf{B}}\alpha^{(n-1)(mp+X_{\mathbf{A}})+mp+(x-1)},\label{encoding:function:2}
\end{IEEEeqnarray}
then it is straightforward to obtain another computation strategy with recovery threshold $K=(n+1)(mp+X_{\mathbf{A}})+X_{\mathbf{B}}-X_{\mathbf{A}}-1$ by following the similar arguments to the one above.

The main result is stated in the following theorem for SSMM.
\begin{Theorem} Consider an SSMM problem where the user is interested in computing the product $\mathbf{C}=\mathbf{AB}$ of the matrices $\mathbf{A}$ and $\mathbf{B}$ with security levels $X_{\mathbf{A}}$ and $X_{\mathbf{B}}$ respectively.
Let $m,n,p\in\mathbb{Z}^{+}$ be the partitioning parameters of the matrices. The proposed computation strategy achieves
\begin{IEEEeqnarray*}{rCl}
\text{Recovery Threshold:}\quad& K,\\
\text{Upload Cost of Sources:}\quad& (U_{\mathbf{A}},U_{\mathbf{B}})=\left(\frac{N}{mp},\frac{N}{np}\right),\\
\text{Download Cost:}\quad&  D=\frac{K}{mn}, \\
\text{Encoding Complexity:}\quad& (\mathcal{C}_{\mathbf{A}},\mathcal{C}_{\mathbf{B}})=\left(\widetilde{\mathcal{O}}\left(\frac{\lambda\xi N(\log N)^{2}}{mp}\right),\widetilde{\mathcal{O}}\left(\frac{\xi\vartheta N(\log N)^2}{np}\right)\right),   \\
\text{Server Computation Complexity:}\quad& \mathcal{C}_s=\mathcal{O}\left(\frac{\lambda\xi\vartheta}{mpn}\right), \\
\text{Decoding Complexity:}\quad& \mathcal{C}_d=\widetilde{\mathcal{O}}\left(\frac{\lambda\vartheta K(\log K)^2}{mn}\right),
\end{IEEEeqnarray*}
where $K=\min\{(m+1)(np+X_{\mathbf{B}})+X_{\mathbf{A}}-X_{\mathbf{B}}-1,(n+1)(mp+X_{\mathbf{A}})+X_{\mathbf{B}}-X_{\mathbf{A}}-1\}$.
\end{Theorem}
\begin{IEEEproof}
Obviously, the recovery threshold $K$ 
can be achieved by the above computation strategies, whose securities, system cost and complexities are proved in Section \ref{proof:strategy}.
\end{IEEEproof}
\begin{Remark}
When $p=1$ and $X_{\mathbf{A}}=X_{\mathbf{B}}$, the degraded encoding functions \eqref{encoding:function:1}, \eqref{encoding:function:2} are the same as that A3S codes \cite{Kakar secure code}.
However, when $n> m$, the degraded functions \eqref{encoded marix:C1}, \eqref{encoded marix:C2} achieve better recovery threshold than A3S codes, see Section \ref{section:comparison} for detailed comparisons.
\end{Remark}

\subsection{Securities, System Cost and Complexities for SSMM}\label{proof:strategy}
Since the two computation strategies presented in Section \ref{construction_SDMM} are similar, we only need to focus on the one with recovery threshold $K=(m+1)(np+X_{\mathbf{B}})+X_{\mathbf{A}}-X_{\mathbf{B}}-1$.

\subsubsection{Security}
Let $v_{x}(\alpha)=\alpha^{{(m-1)(np+X_{\mathbf{B}})+np+(x-1)}}$ (or $v_{x}(\alpha)=\alpha^{{np+(x-1)}}$) for any $x\in[X]$ such that $X\in\mathbb{Z}^{+}$. Clearly, for any $\mathcal{X}=\{i_1,\ldots,i_{X}\}\subseteq [N],|\mathcal{X}|=X$, the following matrix
\begin{IEEEeqnarray*}{c}
\mathbf{V}=
\left[
  \begin{array}{ccc}
    v_1(\alpha_{i_1}) & \ldots & v_{X}(\alpha_{i_1}) \\
    \vdots& \ddots & \vdots \\
    v_1(\alpha_{i_{X}})  & \ldots & v_{X}(\alpha_{i_{X}}) \\
  \end{array}
\right]_{X\times X}
\end{IEEEeqnarray*}
is non-singular, since $\det(\mathbf{V})=(\alpha_{i_1}\ldots\alpha_{i_X})^{(m+1)(np+X_{\mathbf{B}})+np}\cdot\det(\mathbf{V}')$ (or $\det(\mathbf{V})=(\alpha_{i_1}\ldots\alpha_{i_X})^{np}\cdot\det(\mathbf{V}')$),
where $\alpha_{i_1},\ldots,\alpha_{i_X}$ are non-zero distinct elements and $\mathbf{V}'$ is a full-rank Vandermonde matrix with parameters $\alpha_{i_1},\ldots,\alpha_{i_X}$.
Thus, by \eqref{encoded marix:C1}, \eqref{encoded marix:C2} and  Lemma \ref{security proof}, it is straight to prove the securities of data matrices $\mathbf{A}$ and $\mathbf{B}$.
%

\subsubsection{System Cost}
By \eqref{encoded marix:C1} and \eqref{encoded marix:C2}, Source $1$ and Source $2$ share the encoding sub-matrices with sizes of $\frac{\lambda}{m}\times\frac{\xi}{p}$ and $\frac{\xi}{p}\times\frac{\vartheta}{n}$ to each server, respectively. By \eqref{upload cost:B}, the upload cost for  $\mathbf{A}$ and $\mathbf{B}$ are $U_{\mathbf{A}}=\frac{N\times\frac{\lambda}{m}\times\frac{\xi}{p}}{\lambda\times\xi}=\frac{N}{mp},\,
U_{\mathbf{B}}=\frac{N\times\frac{\xi}{p}\times\frac{\vartheta}{n}}{\xi\times\vartheta}=\frac{N}{np}$.
By \eqref{ply:product}, the user downloads a matrix of size $\frac{\lambda}{m}\times\frac{\vartheta}{n}$ from every responsive servers. Thus, according to \eqref{download cost}, the download cost is $D=\frac{K\times\frac{\lambda}{m}\times\frac{\vartheta}{n}}{\lambda\times\vartheta}=\frac{K}{mn}$.

 \subsubsection{Complexity Analysis} The encoding process for matrices $\mathbf{A}$ and $\mathbf{B}$ can be viewed as evaluating a polynomial of degree less than $N$ at $N$ points for $\frac{\lambda\xi}{mp}$ and $\frac{\xi\vartheta}{np}$ times, respectively by \eqref{encoded marix:C1} and \eqref{encoded marix:C2}, decoding requires interpolating a $(K-1)$-th degree polynomial for $\frac{\lambda\vartheta}{mn}$ times  by \eqref{ply:product}.  It is well known \cite{Von} that the evaluation of a $k$-th degree polynomial at $k+1$ arbitrary points can be done in $\widetilde{\mathcal{O}}(k(\log k)^2)$ arithmetic operations, and consequently, its dual problem, interpolation of a $k$-th degree polynomial from $k+1$ arbitrary points can be performed in the same arithmetic operations $\widetilde{\mathcal{O}}(k(\log k)^2)$. Thus, the encoding for matrices $\mathbf{A}$ and $\mathbf{B}$ achieve the complexities $\widetilde{\mathcal{O}}(\frac{\lambda\xi N(\log N)^2}{mp})$ and $\widetilde{\mathcal{O}}(\frac{\xi\vartheta N(\log N)^2}{np})$, respectively, and the decoding achieves the complexity $\widetilde{\mathcal{O}}(\frac{\lambda\vartheta K(\log K)^2}{mn})$. Note from the computation phase that each server is assigned to multiply two coded sub-matrices with sizes of $\frac{\lambda}{m}\times\frac{\xi}{p}$ and $\frac{\xi}{p}\times\frac{\vartheta}{n}$, which requires a complexity of $\mathcal{O}(\frac{\lambda\xi\vartheta}{mpn})$ if the operation of straightforward matrix multiplication is adopted.

\section{Computation Strategies for Secure Multi-Party Batch Matrix Multiplication}\label{section:SMBMM}
In this section, we generalize the computation strategy for SSMM to SMBMM. 
{For clarity, the main parameters used in the SMBMM strategy are listed in Table \ref{tab:parameters}.

\begin{table*}[htbp]
\extrarowheight=4pt
\centering
\caption{Main Parameters Used in SMBMM Strategy}
  \begin{tabular}{|c|c|}
  \hline
  $N$ & number of servers \\
  \hline
  $M$ & number of pairwise matrix products that the user wants to compute \\ 
  \hline
  $\mathbf{A},\mathbf{B}$ & data matrices that are accessed by the two sources \\
  \hline
  $\mathbf{A}\divideontimes\mathbf{B}$ & computation results that the user wants to finish \\
  \hline
    $G$ & number of groups that data matrices are partitioned \\
  \hline
  $L$ & number of matrices in each group \\
  \hline
  $m,p,n$ & partitioning parameters of data matrices \\
  \hline
  $\mathbf{A}^{h,\ell},\mathbf{B}^{h,\ell}$ & the $\ell$-th matrix in the $h$-th group of data matrices $\mathbf{A}$ and $\mathbf{B}$ \\ 
  \hline
  $\mathbf{C}^{h,\ell}$ & product of data matrices $\mathbf{A}^{h,\ell}$ and $\mathbf{B}^{h,\ell}$ \\
  \hline
  $\mathbf{A}_{k,l}^{h,\ell}$ & the $k$-th sub-matrix in the $l$-th column of $\mathbf{A}^{h,\ell}$ \\
  \hline
  $\mathbf{B}_{l,j}^{h,\ell}$ & the $l$-th sub-matrix in the $j$-th column of $\mathbf{B}^{h,\ell}$  \\
  \hline
    $\mathbf{C}_{k,j}^{h,\ell}$ & the $k$-th sub-matrix in the $j$-th column of $\mathbf{C}^{h,\ell}$ \\
  \hline
  $\mathbf{P}^{h,\ell},\mathbf{Q}^{h,\ell}$  &  encoding functions of the partitioning sub-matrices in $\mathbf{A}^{h,\ell}$ and $\mathbf{B}^{h,\ell}$ \\
  \hline
  $\mathbf{H}^{h,\ell}$ & product of the two encoding functions $\mathbf{P}^{h,\ell}$ and $\mathbf{Q}^{h,\ell}$  \\
  \hline
  $X_{\mathbf{A}},X_{\mathbf{B}}$ & number of colluding data-curious servers for matrices $\mathbf{A}$ and $\mathbf{B}$ \\
  \hline
  $\mathbf{Z}_{x}^{\mathbf{A},h},\mathbf{Z}_{x}^{\mathbf{B},h}$ & matrices that are used for ensuring securities of data matrices $\mathbf{A}$ and $\mathbf{B}$ \\
  \hline
    $\widetilde{\mathbf{A}}^{h},\widetilde{\mathbf{B}}^{h}$ & encoding functions of the partitioning sub-matrices in group $h$ for $\mathbf{A}$ and $\mathbf{B}$ \\
  \hline
  $\widetilde{\mathbf{A}}_{i},\widetilde{\mathbf{B}}_{i}$ & encoding matrices that are shared by the sources with server $i$ \\
  \hline
  $\mathcal{S}$ & common randomness across servers \\
  \hline
  $\mathbf{Y}_{i}$ & response of server $i$ \\
  \hline
  $K$ & recovery threshold \\
  \hline
  \end{tabular}
  \label{tab:parameters}
\end{table*}
}





\begin{Lemma}[Cauchy-Vandermonde Matrix with Multiple Poles \cite{Cauchy-Vandermonde matrix}]\label{C-V matrix pole} For any $K,L,r\in\mathbb{Z}^+$ such that $r\leq L<K$,
let $\alpha_1,\alpha_2,\ldots,\alpha_K,d_1,d_2,\ldots,d_r$ be $K+r$ distinct elements from $\mathbb{F}_{q}$ and $L=n_1+n_2+\ldots+n_r$, where $n_1,n_2,\ldots,n_r\in\mathbb{Z}^{+}$, then the following Cauchy-Vandermonde matrix $\mathbf{V}$ is invertible over $\mathbb{F}_q$.
\begin{IEEEeqnarray*}{rCl}\label{C-V matrix1}
\mathbf{V}=
\substack{
\left[
  \begin{array}{ccccccccccc}
    \frac{1}{(d_1-\alpha_1)^{n_1}}& \ldots & \frac{1}{d_1-\alpha_1} &\ldots & \frac{1}{(d_r-\alpha_1)^{n_r}} & \ldots & \frac{1}{d_r-\alpha_1} & 1 & \alpha_1 & \ldots & \alpha_1^{K-L-1} \\
    \frac{1}{(d_1-\alpha_2)^{n_1}}& \ldots & \frac{1}{d_1-\alpha_2} &\ldots & \frac{1}{(d_r-\alpha_2)^{n_r}} & \ldots & \frac{1}{d_r-\alpha_2} & 1 & \alpha_2 & \ldots & \alpha_2^{K-L-1} \\
    \vdots & \ddots & \vdots & \ddots & \vdots & \ddots & \vdots & \vdots & \vdots & \ddots & \vdots \\
    \frac{1}{(d_1-\alpha_K)^{n_1}}& \ldots & \frac{1}{d_1-\alpha_K} &\ldots & \frac{1}{(d_r-\alpha_K)^{n_r}} & \ldots & \frac{1}{d_r-\alpha_K} & 1 & \alpha_K & \ldots & \alpha_K^{K-L-1} \\
  \end{array}
\right]_{K\times K}.\\
\underbrace{~~~~~~~~~~~~~~~~~~~~~~~~~~~~~~~~~~~~~~~~~~~~~~~~~~~~~~~~~~~~}_{\text{\footnotesize{Cauchy part}}}~~
\underbrace{~~~~~~~~~~~~~~~~~~~~~~~~~}_{\text{\footnotesize{Vandermonde part}}}~~~~
}
\end{IEEEeqnarray*}
\end{Lemma}



Similar to \cite{Chen and Jafar,Jia and Jafar}, to better attain the tradeoff with respect to recovery threshold, system cost and complexity, grouping method is employed where each batch of matrices are partitioned into multiple equal-size groups. 
Specifically, given $G,L\in\mathbb{Z}^{+}$ such that $M=GL$ and $L\geq 2$. 
Source $1$ and Source $2$ divide the data matrices $\mathbf{A}$ and $\mathbf{B}$ into $G$ groups of $L$ matrices each, respectively.
Denote by $\mathbf{A}^{h,\ell}=\mathbf{A}^{((h-1)L+\ell)}, \mathbf{B}^{h,\ell}=\mathbf{B}^{((h-1)L+\ell)}$
%
the $\ell$-th matrix of group $h$ for any $\ell\in[L],h\in[G]$.
Consider any partitioning parameters $m>1,n>1,p\geq 1$, the matrices $\mathbf{A}^{h,\ell}$ and $\mathbf{B}^{h,\ell}$ are further divided into $m\times p$ and $p\times n$ equal-size sub-matrices, respectively:
\begin{IEEEeqnarray*}{c}\label{partition:111}
\mathbf{A}^{h,\ell}=
\left[
  \begin{array}{ccc}
    \mathbf{A}_{1,1}^{h,\ell}  & \ldots & \mathbf{A}_{1,p}^{h,\ell} \\
    \vdots  & \ddots & \vdots \\
    \mathbf{A}_{m,1}^{h,\ell}  & \ldots & \mathbf{A}_{m,p}^{h,\ell} \\
  \end{array}
\right],\,
\mathbf{B}^{h,\ell}=
\left[
  \begin{array}{cccc}
    \mathbf{B}_{1,1}^{h,\ell} &  \ldots & \mathbf{B}_{1,n}^{h,\ell} \\
    \vdots & \ddots & \vdots \\
    \mathbf{B}_{p,1}^{h,\ell}  & \ldots & \mathbf{B}_{p,n}^{h,\ell} \\
  \end{array}
\right],\,\forall\,\ell\in[L],h\in[G],
\end{IEEEeqnarray*}
where $\mathbf{A}_{k,l}^{h,\ell}\in\mathbb{F}_q^{\frac{\lambda}{m}\times\frac{\xi}{p}}$ for any $k\in[m],l\in[p]$, and $\mathbf{B}_{l,j}^{h,\ell}\in\mathbb{F}_q^{\frac{\xi}{p}\times\frac{\vartheta}{n}}$ for any $l\in[p],j\in[n]$.

Then, the objective of the user is to securely compute the desired matrix products $(\mathbf{C}^{h,1},\ldots,\mathbf{C}^{h,L})=(\mathbf{A}^{h,1}\mathbf{B}^{h,1},\ldots,\mathbf{A}^{h,L}\mathbf{B}^{h,L})$ for each group $h\in[G]$, where
\begin{IEEEeqnarray}{c}\label{desred:product2}
\mathbf{C}^{h,\ell}=\mathbf{A}^{h,\ell}\mathbf{B}^{h,\ell}=
\left[
  \begin{array}{ccc}
    \mathbf{C}_{1,1}^{h,\ell} & \ldots & \mathbf{C}_{1,n}^{h,\ell} \\
    \vdots & \ddots & \vdots \\
    \mathbf{C}_{m,1}^{h,\ell} & \ldots & \mathbf{C}_{m,n}^{h,\ell} \\
  \end{array}
\right], \quad\forall\, \ell\in[L],h\in[G]
\end{IEEEeqnarray}
with $\mathbf{C}_{k,j}^{h,\ell}=\sum_{l=1}^{p}\mathbf{A}_{k,l}^{h,\ell}\mathbf{B}_{l,j}^{h,\ell}$ for any $k\in[m],j\in[n]$.
For this purpose, we use the functions $\widetilde{\mathbf{A}}^{h}(\alpha)$ and $\widetilde{\mathbf{B}}^h(\alpha)$ to encode all the partitioning sub-matrices $\{\mathbf{A}_{k,l}^{h,\ell}:k\in[m],l\in[p],\ell\in[L]\}$ and $\{\mathbf{B}_{l,j}^{h,\ell}:l\in[p],j\in[n],\ell\in[L]\}$ in group $h\in[G]$, respectively. Similar to C1-C3, the two encoding functions are constructed such that the following criteria are satisfied:
\begin{itemize}
\item[C1$'$] The function $\widetilde{\mathbf{A}}^{h}(\alpha)\cdot \widetilde{\mathbf{B}}^{h}(\alpha)$ contains the desired sub-products $\mathbf{C}_{k,j}^{h,\ell},k\in[m],j\in[n],\ell\in[L]$. Meanwhile, the dimensions corresponding to these desired terms are independent from each other and are  from the dimensions corresponding to all the remaining terms appearing in $\widetilde{\mathbf{A}}^{h}(\alpha)\cdot \widetilde{\mathbf{B}}^{h}(\alpha)$ as well. Moreover, in the response function $\sum_{h\in[G]}\widetilde{\mathbf{A}}^{h}(\alpha)\cdot \widetilde{\mathbf{B}}^{h}(\alpha)$, these desired sub-products must be separated with each other across all the groups.  This guarantees decodability.
\item[C2$'$] Except for these desired terms in $\widetilde{\mathbf{A}}^{h}(\alpha)\cdot \widetilde{\mathbf{B}}^{h}(\alpha)$, the remaining undesired components should be aligned as much as possible. This will allow us to minimize the recovery threshold.

\item[C3$'$] The functions $\widetilde{\mathbf{A}}^{h}(\alpha)$ and $\widetilde{\mathbf{B}}^h(\alpha)$ are encoded by mixing the partitioning sub-matrices with random noise, and the terms corresponding to the noises occupy independent dimensions. This ensures the data security from the servers by Lemma \ref{security proof}.
\item[C4$'$] In server responses, the idea of noise alignment \cite{Zhao} is employed in the same structure as $\sum_{h\in[G]}\widetilde{\mathbf{A}}^{h}(\alpha)\cdot \widetilde{\mathbf{B}}^{h}(\alpha)$ such that the desired terms are perfectly reserved and the other terms are aligned with random noises. This ensures the data security for the user beyond the products $\mathbf{A}\ast\mathbf{B}$.
\end{itemize}

 In our SMBMM strategy, C1$'$ and C2$'$ are achieved by employing Cross Subspace Alignment (CSA) idea \cite{CSA_PIR,Jia and Jafar} that facilitates a form of interference alignment (i.e., Cauchy-Vandermonde structure) to separate the desired terms and interference in $\widetilde{\mathbf{A}}^{h}(\alpha)\cdot\widetilde{\mathbf{B}}^{h}(\alpha)$, such that the desired terms appear along the dimensions corresponding to the \emph{Cauchy part} and interference is aligned as much as possible along the dimensions corresponding to the \emph{Vandermonde part}.  Moreover, in the response function $\sum_{h\in[G]}\widetilde{\mathbf{A}}^{h}(\alpha)\cdot \widetilde{\mathbf{B}}^{h}(\alpha)$, interference can be further aligned when summing the product $\widetilde{\mathbf{A}}^{h}(\alpha)\cdot \widetilde{\mathbf{B}}^{h}(\alpha)$ over $h=1,\ldots,G$, since the interference appear along the Vandermonde part across all the groups. The criteria  C3$'$ and C4$'$ are also easy to be satisfied by such Cauchy-Vandermonde structure due to Lemma \ref{C-V matrix pole}. 

\subsection{Illustrative Example for SMBMM}\label{SMBMM:example2}

We will extend the example in Section \ref{example:1} to the problem of SMBMM with parameters $m=2,p=3,n=2,X_{\mathbf{A}}=2,X_{\mathbf{B}}=3,M=4$. The example is presented as much detail as possible to illustrate the essential ingredients of the SMBMM strategy.

The strategy splits the $M=GL=4$ instances of data matrices $\mathbf{A}=(\mathbf{A}^{1,1},\mathbf{A}^{1,2},\mathbf{A}^{2,1},\mathbf{A}^{2,2})$ and $\mathbf{B}=(\mathbf{B}^{1,1},\mathbf{B}^{1,2},\mathbf{B}^{2,1},\mathbf{B}^{2,2})$ into $G=2$ group respectively, each containing $L=2$ matrices, i.e., $(\mathbf{A}^{1,1},\mathbf{A}^{1,2})$ and $(\mathbf{A}^{2,1},\mathbf{A}^{2,2})$ (resp. $(\mathbf{B}^{1,1},\mathbf{B}^{1,2})$ and $(\mathbf{B}^{2,1},\mathbf{B}^{2,2})$).

Then, partition $\mathbf{A}^{h,\ell}$ and $\mathbf{B}^{h,\ell}$ as \eqref{desired product:exam222} and \eqref{desired product:exam}, and rewrite them
as $\mathbf{A}^{h,\ell}=\big[\mathbf{A}_{k,l}^{h,\ell}\big]_{k\in [2],l\in [3]}$ and $\mathbf{B}^{h,\ell}=\big[\mathbf{B}_{l,j}^{h,\ell}\big]_{l\in [3],j\in [2]}$ for $\ell\in [2],h\in[2]$, respectively.
Denote the desired product $\mathbf{C}^{h,\ell}=\mathbf{A}^{h,\ell}\mathbf{B}^{h,\ell}=\big[\mathbf{C}_{k,j}^{h,\ell}\big]_{k\in [2],j\in [2]}$,
where $\mathbf{C}_{k,j}^{h,\ell}=\mathbf{A}_{k,1}^{h,\ell}\mathbf{B}_{1,j}^{h,\ell}+\mathbf{A}_{k,2}^{h,\ell}\mathbf{B}_{2,j}^{h,\ell}+\mathbf{A}_{k,3}^{h,\ell}\mathbf{B}_{3,j}^{h,\ell}$.

To obtain the two encoding functions $\widetilde{\mathbf{A}}^{h}(\alpha)$ and $\widetilde{\mathbf{B}}^h(\alpha)$ satisfying  C1$'$-C4$'$, we first construct the sub-functions $\mathbf{P}^{h,\ell}(\alpha)$ and $\mathbf{Q}^{h,\ell}(\alpha)$ by encoding the sub-matrices in $\mathbf{A}^{h,\ell}$ and $\mathbf{B}^{h,\ell}$ respectively, and then use  Cauchy and Vandermonde parameters to combine these sub-functions.

Let $f_{1,1},f_{1,2},f_{2,1},f_{2,2}$ be $4$ distinct elements from $\mathbb{F}_q$.
Then, according to the construction of SSMM \eqref{sharing:1}-\eqref{sharing:2}, define
\begin{IEEEeqnarray}{rCl}
\mathbf{P}^{h,1}(\alpha)&=&\sum\limits_{k=1}^{2}\left(\sum\limits_{l=1}^{3}\mathbf{A}_{k,l}^{h,1}(f_{h,1}-\alpha)^{l-1}\right)(f_{h,1}-\alpha)^{9(k-1) }+\sum\limits_{x=1}^{2}\mathbf{Z}_{x}^{\mathbf{A},h}(f_{h,1}-\alpha)^{14+x}, \label{example:Ph1}\\
\mathbf{Q}^{h,1}(\alpha)&=&\sum\limits_{j=1}^{2}\left(\sum\limits_{l=1}^{3}\mathbf{B}_{l,j}^{h,1}(f_{h,1}-\alpha)^{3-l}\right)(f_{h,1}-\alpha)^{3(j-1)}+\sum\limits_{x=1}^{3}\mathbf{Z}_{x}^{\mathbf{B},h}(f_{h,1}-\alpha)^{5+x} \label{example:Qh1}
\end{IEEEeqnarray}
for $\mathbf{A}^{h,1}$ and $\mathbf{B}^{h,1}$, where $\mathbf{Z}_{1}^{\mathbf{A},h},\mathbf{Z}_{2}^{\mathbf{A},h},\mathbf{Z}_{1}^{\mathbf{B},h},\mathbf{Z}_{2}^{\mathbf{B},h}$ are corresponding random matrices for each $h\in[2]$. While for $\mathbf{A}^{h,2}$ and $\mathbf{B}^{h,2}$, following the construction of  entangled polynomials \cite{EP code}, let
\begin{IEEEeqnarray*}{rCll}
\mathbf{P}^{h,2}(\alpha)&=&\sum\limits_{k=1}^{2}\left(\sum\limits_{l=1}^{3}\mathbf{A}_{k,l}^{h,2}(f_{h,2}-\alpha)^{l-1}\right){(f_{h,2}-\alpha)^{6(k-1)}},\\
\mathbf{Q}^{h,2}(\alpha)&=&\sum\limits_{j=1}^{2}\left(\sum\limits_{l=1}^{3}\mathbf{B}_{l,j}^{h,2}(f_{h,2}-\alpha)^{3-l}\right){(f_{h,2}-\alpha)^{3(j-1)}}.
\end{IEEEeqnarray*}
Notably, compared to the constructions of SSMM and entangled polynomials, the shifted versions of $\alpha$ in distinct parameters $f_{h,\ell}$ across all $h\in[2],\ell\in[2]$ are used to distinguish the batch of $M=GL=4$ desired products.
Moreover, let
\begin{IEEEeqnarray}{l}\label{desired:examp}
\mathbf{H}^{h,\ell}(\alpha)=\mathbf{P}^{h,\ell}(\alpha)\cdot\mathbf{Q}^{h,\ell}(\alpha)=
\left\{\begin{array}{@{}ll}
\sum\limits_{r=0}^{24}\mathbf{H}_r^{h,\ell}(f_{h,\ell}-\alpha)^r, &\quad \ell=1\\
\sum\limits_{r=0}^{13}\mathbf{H}_r^{h,\ell}(f_{h,\ell}-\alpha)^r, &\quad \ell=2
\end{array}
\right.,
\end{IEEEeqnarray}
where $\mathbf{H}_r^{h,\ell}$ is the coefficient of $(f_{h,\ell}-\alpha)^r$, specially,
\begin{IEEEeqnarray}{rClrClrClrCl}
\mathbf{H}_{2}^{h,1}&=&\mathbf{C}_{1,1}^{h,1}, \quad \mathbf{H}_{5}^{h,1}&=&\mathbf{C}_{1,2}^{h,1},  \quad \mathbf{H}_{11}^{h,1}&=&\mathbf{C}_{2,1}^{h,1}, \quad \mathbf{H}_{14}^{h,1}&=&\mathbf{C}_{2,2}^{h,1},  \label{desired:coeff:1}\\
\mathbf{H}_{2}^{h,2}&=&\mathbf{C}_{1,1}^{h,2}, \quad \mathbf{H}_{5}^{h,2}&=&\mathbf{C}_{1,2}^{h,2},  \quad \mathbf{H}_{8}^{h,2}&=&\mathbf{C}_{2,1}^{h,2},  \quad \mathbf{H}_{11}^{h,2}&=&\mathbf{C}_{2,2}^{h,2}.\label{desired:coeff:44}
\end{IEEEeqnarray}

Obviously, by \eqref{desired:coeff:1} and \eqref{desired:coeff:44}, the desired products $\mathbf{C}^{h,1},\mathbf{C}^{h,2}$ ($h\in[2]$) can be recovered if the coefficients
$\{\mathbf{H}_r^{h,1}:r\in[0:14],h\in[2]\}$ and $\{\mathbf{H}_r^{h,2}:r\in[0:11],h\in[2]\}$ are obtained.
The secure encoding functions for the matrices of each group are constructed as
\begin{IEEEeqnarray*}{rCll}
\widetilde{\mathbf{A}}^{h}(\alpha)&=&\Delta^{h}(\alpha)\left( \frac{1}{(f_{h,1}-\alpha)^{15}}\mathbf{P}^{h,1}(\alpha)+\frac{1}{(f_{h,2}-\alpha)^{12}}\mathbf{P}^{h,2}(\alpha) \right),  \label{encdoing:examp1}\\
\widetilde{\mathbf{B}}^{h}(\alpha)&=&
\frac{1}{(f_{h,1}-\alpha)^{15}}\mathbf{Q}^{h,1}(\alpha)+\frac{1}{(f_{h,2}-\alpha)^{12}}\mathbf{Q}^{h,2}(\alpha), \label{encdoing:examp2}
\end{IEEEeqnarray*}
where $\Delta^{h}(\alpha)=(f_{h,1}-\alpha)^{15}\cdot(f_{h,2}-\alpha)^{12}$ for $h=1,2$ is used for aligning the interference from the undesired cross products $\mathbf{P}^{h,1}\mathbf{Q}^{h,2},\mathbf{P}^{h,2}\mathbf{Q}^{h,1}$ within dimensions corresponding to the Vandermonde part, as shown in the third term in \eqref{expanding form222}. Remarkably, the criterion C3$'$ is satisfied due to \eqref{example:Ph1} and \eqref{example:Qh1}.
Further, by expanding $\widetilde{\mathbf{A}}^{h}(\alpha)\cdot\widetilde{\mathbf{B}}^{h}(\alpha)$, it is straight to observe that the two functions satisfy the criterion C2$'$:
\begin{IEEEeqnarray}{rCl}
&&\widetilde{\mathbf{A}}^{h}(\alpha)\cdot\widetilde{\mathbf{B}}^{h}(\alpha) \notag\\
&=&\frac{(f_{h,2}-\alpha)^{12}}{(f_{h,1}-\alpha)^{15}}\mathbf{P}^{h,1}(\alpha)\mathbf{Q}^{h,1}(\alpha)+\frac{(f_{h,1}-\alpha)^{15}}{(f_{h,2}-\alpha)^{12}}\mathbf{P}^{h,2}(\alpha)\mathbf{Q}^{h,2}(\alpha) +\mathbf{P}^{h,1}(\alpha)\mathbf{Q}^{h,2}(\alpha)+\mathbf{P}^{h,2}(\alpha)\mathbf{Q}^{h,1}(\alpha)\notag \\
&=&\sum\limits_{r=0}^{14}\sum\limits_{s=r+1}^{15}\frac{c_{15-s}^{h,1}}{(f_{h,1}-\alpha)^{s-r}}\mathbf{H}_r^{h,1}+
\sum\limits_{r=0}^{11}\sum\limits_{s=r+1}^{12}\frac{c_{12-s}^{h,2}}{(f_{h,2}-\alpha)^{s-r}}\mathbf{H}_r^{h,2}
+\sum\limits_{r=0}^{21}\mathbf{U}_r^{h}\alpha^r, \label{expanding form222}
\end{IEEEeqnarray}
where $c_{13}^{h,1}=c_{14}^{h,1}=0$, and the other coefficients $c_s^{h,\ell}$ are determined by expanding
\begin{IEEEeqnarray}{c}
\big((f_{h,1}-\alpha)+(f_{h,2}-f_{h,1})\big)^{12}=\sum\limits_{s=0}^{14}c_s^{h,1}(f_{h,1}-\alpha)^{s},\label{coefficients:123} \\
\big((f_{h,2}-\alpha)+(f_{h,1}-f_{h,2})\big)^{15}=\sum\limits_{s=0}^{15}c_s^{h,2}(f_{h,2}-\alpha)^{s},\label{coefficients:234}
\end{IEEEeqnarray}
and $\mathbf{U}_{r}^{h}$ is the interference corresponding to the Vandermonde term $\alpha^r$ for any $r\in[0:21]$, which represents various combinations of sub-matrix products and whose exact forms are unimportant.
That is, in the computation \eqref{expanding form222}, the desired coefficients $\{\mathbf{H}_r^{h,1}:r\in[0:14]\}\cup\{\mathbf{H}_r^{h,2}:r\in[0:11]\}$ appear along the Cauchy terms $\big\{\sum_{s=r+1}^{15}\frac{c_{15-s}^{h,1}}{(f_{h,1}-\alpha)^{s-r}}:r\in[0:14]\big\}\cup\big\{\sum_{s=r+1}^{12}\frac{c_{12-s}^{h,2}}{(f_{h,2}-\alpha)^{s-r}}:r\in[0:11]\big\}$, and interference align along the Vandermonde terms $\{\alpha^{r}:r\in[0:21]\}$ of $22$ dimensions for any $h\in[2]$.

Let $\alpha_1,\ldots,\alpha_N$ be $N$ distinct elements from $\mathbb{F}_q\backslash\{f_{1,1},f_{1,2},f_{2,1},f_{2,2}\}$. In the sharing phase, Source $1$ and Source $2$ share the encoding sub-matrices $\widetilde{\mathbf{A}}_{i}=(\widetilde{\mathbf{A}}^{1}(\alpha_{i}),\widetilde{\mathbf{A}}^{2}(\alpha_{i}))$ and $\widetilde{\mathbf{B}}_{i}=(\widetilde{\mathbf{B}}^{1}(\alpha_{i}),\widetilde{\mathbf{B}}^{2}(\alpha_{i}))$ with server $i$, respectively.
Denote by $\Gamma=\{2,5,11,14\},\,\Lambda=\{2,5,8,11\}$ the indices used for recovering desired products by \eqref{desired:coeff:1} and \eqref{desired:coeff:44}.
To ensure the data security for the user beyond the desired products, let
\begin{IEEEeqnarray}{rCl}\label{shared variables:1}
\mathcal{S}=\{\mathbf{Z}_{r}^{h,1}:r\in[0:14]\backslash\Gamma,h\in[2]\}\cup\{\mathbf{Z}_{r}^{h,2}:r\in[0:11]\backslash\Lambda,h\in[2]\}\cup\{\mathbf{Z}_r:r\in[0:21]\}\IEEEeqnarraynumspace
\end{IEEEeqnarray}
be the shared random matrices across servers, chosen independently and uniformly from all matrices over $\mathbb{F}_q^{\frac{\lambda}{2}\times\frac{\vartheta}{2}}$. 
In addition, set
\begin{IEEEeqnarray}{rCl}
\mathbf{Z}_{r}^{h,1}&=&[\mathbf{0}]_{\frac{\lambda}{2}\times\frac{\vartheta}{2}},\quad r\in\Gamma,h\in[2], \label{zeros:example}\\
\mathbf{Z}_{r}^{h,2}&=&[\mathbf{0}]_{\frac{\lambda}{2}\times\frac{\vartheta}{2}},\quad r\in\Lambda,h\in[2].\label{set:01}
\end{IEEEeqnarray}

Following \eqref{expanding form222}, define a noise polynomial
\begin{IEEEeqnarray*}{rCl}
\mathbf{S}(\alpha)&=&\sum\limits_{h=1}^{2}\sum\limits_{r=0}^{14}\sum\limits_{s=r+1}^{15}\frac{c_{15-s}^{h,1}}{(f_{h,1}-\alpha)^{s-r}}\mathbf{Z}_r^{h,1}+
\sum\limits_{h=1}^{2}\sum\limits_{r=0}^{11}\sum\limits_{s=r+1}^{12}\frac{c_{12-s}^{h,2}}{(f_{h,2}-\alpha)^{s-r}}\mathbf{Z}_r^{h,2}
+\sum\limits_{r=0}^{21}\mathbf{Z}_r\alpha^r.
\end{IEEEeqnarray*}

Then, server $i$ computes the response from all the received massages:
\begin{IEEEeqnarray}{rCl}
\mathbf{Y}_{i}&=&\sum\limits_{h=1}^{2}\widetilde{\mathbf{A}}^{h}(\alpha_i)\cdot\widetilde{\mathbf{B}}^{h}(\alpha_i)+\mathbf{S}(\alpha_i) =\sum\limits_{h=1}^{2}\sum\limits_{r=0}^{14}\sum\limits_{s=r+1}^{15}\frac{c_{15-s}^{h,1}}{(f_{h,1}-\alpha_i)^{s-r}}(\mathbf{H}_r^{h,1}+\mathbf{Z}_r^{h,1})\notag\\
&&\quad\quad\quad\quad\quad+\sum\limits_{h=1}^{2}\sum\limits_{r=0}^{11}\sum\limits_{s=r+1}^{12}\frac{c_{12-s}^{h,2}}{(f_{h,2}-\alpha_i)^{s-r}}(\mathbf{H}_r^{h,2}+\mathbf{Z}_r^{h,2})+\sum\limits_{r=0}^{21}(\mathbf{U}_r+\mathbf{Z}_r)\alpha_i^r, \label{response:example2}
\end{IEEEeqnarray}
where $\mathbf{U}_r=\mathbf{U}_r^{1}+\mathbf{U}_r^{2}$ for any $r\in[0:21]$. Thus, the answers from any $K=76$ servers, indexed by $i_1,i_2,\ldots,i_K$, 
can be denoted in the matrix form as follow.
\begin{figure}[htp]
\begin{IEEEeqnarray}{c}
\left[
\begin{array}{@{}c@{}}
\mathbf{Y}_{i_1} \\
\mathbf{Y}_{i_2} \\
\vdots \\
\mathbf{Y}_{i_K}
\end{array}
\right]=\big((
\underbrace{
\left[
\begin{array}{ccc;{2pt/2pt}c;{2pt/2pt}ccc;{2pt/2pt}ccccc}
\frac{1}{(f_{1,1}-\alpha_{i_1})^{15}} & \ldots & \frac{1}{f_{1,1}-\alpha_{i_1}} & \ldots & \frac{1}{(f_{2,2}-\alpha_{i_1})^{12}} & \ldots & \frac{1}{f_{2,2}-\alpha_{i_1}} & 1 & \alpha_{i_1}  & \ldots & \alpha_{i_1}^{21} \\
\frac{1}{(f_{1,1}-\alpha_{i_2})^{15}} & \ldots & \frac{1}{f_{1,1}-\alpha_{i_2}} & \ldots & \frac{1}{(f_{2,2}-\alpha_{i_2})^{12}} & \ldots & \frac{1}{f_{2,2}-\alpha_{i_2}} & 1  & \alpha_{i_2}  & \ldots & \alpha_{i_2}^{21} \\
\vdots & \vdots   & \vdots & \vdots & \vdots & \vdots & \vdots & \vdots & \vdots  & \vdots & \vdots \\
\frac{1}{(f_{1,1}-\alpha_{i_K})^{15}} & \ldots  & \frac{1}{f_{1,1}-\alpha_{i_K}} & \ldots & \frac{1}{(f_{2,2}-\alpha_{i_K})^{12}}  & \ldots & \frac{1}{f_{2,2}-\alpha_{i_K}} & 1 & \alpha_{i_K}  & \ldots & \alpha_{i_K}^{21}
\end{array}
\right]
}_{\mathbf{V}_1} \notag\\
\underbrace{
\left[
\begin{array}{c;{2pt/2pt}c;{2pt/2pt}c;{2pt/2pt}c}
\textbf{T}(c_{0}^{1,1},\ldots,c_{14}^{1,1})&&& \\ \hdashline[2pt/2pt]
&\ddots & &\\\hdashline[2pt/2pt]
& & \textbf{T}(c_{0}^{2,2},\ldots,c_{11}^{2,2})& \\\hdashline[2pt/2pt]
& & & \mathbf{I}_{22} \\
\end{array}
\right]
}_{\mathbf{V}_2})
\otimes\mathbf{I}_{\lambda/2}\big)
\left[
\begin{array}{@{}c@{}}
\mathbf{H}_0^{1,1}+\mathbf{Z}_0^{1,1} \\
\vdots \\
\mathbf{H}_{14}^{1,1}+\mathbf{Z}_{14}^{1,1} \\\hdashline[2pt/2pt]
\vdots \\\hdashline[2pt/2pt]
\mathbf{H}_0^{2,2}+\mathbf{Z}_0^{2,2} \\
\vdots \\
\mathbf{H}_{11}^{2,2}+\mathbf{Z}_{11}^{2,2} \\\hdashline[2pt/2pt]
\mathbf{U}_0+\mathbf{Z}_0 \\
\mathbf{U}_1+\mathbf{Z}_1 \\
\vdots \\
\mathbf{U}_{21}+\mathbf{Z}_{21} \\
\end{array}
\right]. \label{answers:multi party2222}
\end{IEEEeqnarray}
\end{figure}

Recall that $f_{1,1}, f_{1,2},f_{2,1},f_{2,2}$ are distinct. Thus all the constants $c_{0}^{1,1},c_{0}^{1,2},c_{0}^{2,1},c_{0}^{2,2}$ take non-zero values by \eqref{coefficients:123} and \eqref{coefficients:234}.
Following from Lemma \ref{C-V matrix pole} and the fact that Kronecker product of non-singular matrices is non-singular, the matrix $\mathbf{V}_1\mathbf{V}_2\otimes\mathbf{I}_{\lambda/m}$ is invertible and thus the user is able to decode $\{\mathbf{H}_r^{h,1}+\mathbf{Z}_{r}^{h,1}: r\in[0:14],h\in[2]\}$ and $\{\mathbf{H}_r^{h,2}+\mathbf{Z}_{r}^{h,2}:r\in[0:11],h\in[2]\}$.
%
By \eqref{zeros:example} and \eqref{set:01}, we further obtain $\{\mathbf{H}_{r}^{h,1},r\in\Gamma,h\in[2]\}$ and $\{\mathbf{H}_{r}^{h,2},r\in\Lambda,h\in[2]\}$.
Thus, the desired product $\mathbf{A}\divideontimes\mathbf{B}=(\mathbf{A}^{1,1}\mathbf{B}^{1,1},\ldots,\mathbf{A}^{2,2}\mathbf{B}^{2,2})$ can be recovered by \eqref{desired:coeff:1} and \eqref{desired:coeff:44}.
The computation strategy achieves the recovery threshold $K=76$. Moreover, the invertible matrix $\mathbf{V}_1\mathbf{V}_2\otimes\mathbf{I}_{\lambda/m}$ means that the desired terms are separated from each other and all the remaining undesired components, i.e., the criterion C1$'$ is satisfied.

Obviously, no information about the data matrices is leaked to the user more than the result of the computation since the noise polynomial $\mathbf{S}(\alpha)$ is structured in the same manner as $\sum_{h=1}^{2}\widetilde{\mathbf{A}}^{h}(\alpha)\cdot\widetilde{\mathbf{B}}^{h}(\alpha)$,i.e., the criterion C4$'$ is satisfied.
By \eqref{shared variables:1}, the servers share $60$ random matrices of dimensions $\frac{\lambda}{2}\times\frac{\vartheta}{2}$ and thus the amount of common randomness $\rho$ is $\frac{15}{4}$.  
\begin{Remark}\label{reducing random}
By expanding \eqref{expanding form222}, we notice that the term $\mathbf{U}_{21}$ with the highest power of $\alpha$ has the form of
\begin{IEEEeqnarray*}{c}
\mathbf{U}_{21}=\sum\limits_{h=1}^{2}\mathbf{U}_{21}^{h}=\sum\limits_{h=1}^{2}(-\mathbf{H}_{24}^{h,1}-\mathbf{Z}_{2}^{\mathbf{A},h}\mathbf{B}_{1,2}^{h,2})=\sum\limits_{h=1}^{2}(-\mathbf{Z}_{2}^{\mathbf{A},h}\mathbf{Z}_{3}^{\mathbf{B},h}-\mathbf{Z}_{2}^{\mathbf{A},h}\mathbf{B}_{1,2}^{h,2}),
\end{IEEEeqnarray*}
where $\mathbf{Z}_{2}^{\mathbf{A},h},\mathbf{Z}_{3}^{\mathbf{B},h}$ are the random matrices generated by sources but unknown to the user for all $h\in[2]$. Thus, $\mathbf{U}_{21}$ leaks nothing about the data matrices to the user and it is unnecessary to use the aligned noise $\mathbf{Z}_{21}$ to scramble the term $\mathbf{U}_{21}$. That is, the random noises that are used for protecting data security from the servers can be exploited to further reduce the number of random matrices shared at the servers from $60$ to $59$, and thus correspondingly the amount of common randomness $\rho$ is reduced to $\frac{59}{16}$.
\end{Remark}

\subsection{General Computation Strategy for SMBMM}\label{General SMBMM}
In this subsection, we present the general construction for the computation strategy of SMBMM.
The strategy  outperforms \cite{Chen and Jafar} when the degraded security level $X_{\mathbf{A}}=X_{\mathbf{B}}$ is less than a threshold, see Table \ref{tab:comparative33} for comparison.
In general, both \cite{Chen and Jafar} and this current work generalize the problem of bath matrix multiplication to the setup of SMBMM, based on the ideas of CSA codes \cite{Jia and Jafar}, Shamir's secret sharing scheme \cite{Shamir} and noise alignment \cite{Zhao}. Particularly, we carefully design the encoding functions $\mathbf{A}^{h,\ell}$ and $\mathbf{B}^{h,\ell}$  to align the interference in $\mathbf{A}^{h,\ell}\cdot\mathbf{B}^{h,\ell}$ to the most extent,  which results in the  reduction of recovery threshold.

In the following, we start by describing the constructions of the functions $\widetilde{\mathbf{A}}^{h}(\alpha)$ and $\widetilde{\mathbf{B}}^h(\alpha)$, based on CSA idea and the criteria C1$'$-C4$'$.
For each group $h\in[G]$,  Source $1$ generates $X_{\mathbf{A}}$ random matrices $\mathbf{Z}_{1}^{\mathbf{A},h},\ldots,\mathbf{Z}_{X_{\mathbf{A}}}^{\mathbf{A},h}$, chosen independently and uniformly from $\mathbb{F}_q^{\frac{\lambda}{m}\times\frac{\xi}{p}}$. Similarly, Source $2$ generates $X_{\mathbf{B}}$ random matrices $\mathbf{Z}_{1}^{\mathbf{B},h},\ldots,\mathbf{Z}_{X_{\mathbf{B}}}^{\mathbf{B},h}$ from $\mathbb{F}_q^{\frac{\xi}{p}\times\frac{\vartheta}{n}}$.
Let $f_{1,1},\ldots,f_{G,L}$ be $GL$ distinct elements on $\mathbb{F}_q$.
We first create the  polynomial functions $\mathbf{P}^{h,\ell}(\alpha)$ and $\mathbf{Q}^{h,\ell}(\alpha)$ to encode the sub-matrices in $\mathbf{A}^{h,\ell}$ and $\mathbf{B}^{h,\ell}$ respectively, for each $\ell\in[L]$ as follows.

For $\ell=1,h\in[G]$, $\mathbf{P}^{h,1}(\alpha)$ and $\mathbf{Q}^{h,1}(\alpha)$ are given by a shifted version of  \eqref{encoded marix:C1} and \eqref{encoded marix:C2}:
\begin{IEEEeqnarray}{rCl}
\mathbf{P}^{h,1}(\alpha)&=&\sum\limits_{k=1}^{m}\left(\sum\limits_{l=1}^{p}\mathbf{A}_{k,l}^{h,1}(f_{h,1}-\alpha)^{l-1}\right)(f_{h,1}-\alpha)^{(k-1)(np+X_{\mathbf{B}})} \notag\\
&&\quad\quad\quad\quad\quad\quad\quad\quad\quad\quad\quad\quad\quad+\sum\limits_{x=1}^{X_{\mathbf{A}}}\mathbf{Z}_{x}^{\mathbf{A},h}(f_{h,1}-\alpha)^{{(m-1)(np+X_{\mathbf{B}})+np+(x-1)}},
\label{encoding function:h,1:1}\\
\mathbf{Q}^{h,1}(\alpha)&=&\sum\limits_{j=1}^{n}\left(\sum\limits_{l=1}^{p}\mathbf{B}_{l,j}^{h,1}(f_{h,1}-\alpha)^{p-l}\right)(f_{h,1}-\alpha)^{(j-1)p}+\sum\limits_{x=1}^{X_{\mathbf{B}}}\mathbf{Z}_{x}^{\mathbf{B},h}(f_{h,1}-\alpha)^{np+(x-1)}.
\label{encoding function:h,1:2}
\end{IEEEeqnarray}
For $\ell\in[2:L],h\in[G]$,  $\mathbf{P}^{h,\ell}(\alpha)$ and $\mathbf{Q}^{h,\ell}(\alpha)$ are created using the shifted version of Entangled Polynomials in \cite{EP code}:
\begin{IEEEeqnarray}{rCll}
\mathbf{P}^{h,\ell}(\alpha)&=&\sum\limits_{k=1}^{m}\left(\sum\limits_{l=1}^{p}\mathbf{A}_{k,l}^{h,\ell}(f_{h,\ell}-\alpha)^{l-1}\right){(f_{h,\ell}-\alpha)^{(k-1)np}},&\quad\forall\,\ell\in[2:L], h\in[G], \label{P:alp}\\
\mathbf{Q}^{h,\ell}(\alpha)&=&\sum\limits_{j=1}^{n}\left(\sum\limits_{l=1}^{p}\mathbf{B}_{l,j}^{h,\ell}(f_{h,\ell}-\alpha)^{p-l}\right){(f_{h,\ell}-\alpha)^{(j-1)p}} ,&\quad\forall\,\ell\in[2:L],h\in[G].\label{Q:alp}
\end{IEEEeqnarray}

Notably, in the original constructions of SSMM and EP codes, the encoding functions $\mathbf{P}^{h,\ell}(\alpha)$ and $\mathbf{Q}^{h,\ell}(\alpha)$ are polynomials of $\alpha$, however in the construction of SMBMM, they are created by using the shifted version of $\alpha$ in distinct parameters $f_{h,\ell}$ across all $\ell\in[L],h\in[G]$.
This is because, instead of computing individual matrix product, the user uses distinct shifted parameters $f_{h,\ell}$ in the encoding polynomials to distinguish the batch of $M=GL$ matrix products $\{\mathbf{A}^{h,\ell}\mathbf{B}^{h,\ell}\}_{\ell\in[L],h\in[G]}$ with each other for SMBMM.

Similar to \eqref{delta:def}-\eqref{desired:coeffi}, denote the product of $\mathbf{P}^{h,\ell}(\alpha)$ and $\mathbf{Q}^{h,\ell}(\alpha)$ by
\begin{IEEEeqnarray}{l}
\mathbf{H}^{h,\ell}(\alpha)=\mathbf{P}^{h,\ell}(\alpha)\cdot\mathbf{Q}^{h,\ell}(\alpha)=
\left\{\begin{array}{@{}ll}
\sum\limits_{r=0}^{\delta}\mathbf{H}_r^{h,1}(f_{h,1}-\alpha)^r,  &\quad \ell=1, h\in[G]  \\
\sum\limits_{r=0}^{mpn+p-2}\mathbf{H}_r^{h,\ell}(f_{h,\ell}-\alpha)^r,&\quad \ell\in[2:L], h\in[G]
\end{array}
\right.,\label{H:alp}
\end{IEEEeqnarray}
where $\mathbf{H}_{r}^{h,\ell}$ is the coefficient of $(f_{h,\ell}-\alpha)^r$ for all $\ell\in[L],h\in[G]$ and $\delta=(m+1)(np+X_{\mathbf{B}})+X_{\mathbf{A}}-X_{\mathbf{B}}-2$ is the degree of polynomial $\mathbf{P}^{h,1}(\alpha)\cdot\mathbf{Q}^{h,1}(\alpha)$ for all $h\in[G]$. In particular, the coefficients satisfy
\begin{IEEEeqnarray}{rCll}
\mathbf{H}_{(k-1)(np+X_{\mathbf{B}})+jp-1}^{h,1}&=&\sum\limits_{l=1}^{p}\mathbf{A}_{k,l}^{h,1}\mathbf{B}_{l,j}^{h,1}=\mathbf{C}_{k,j}^{h,1},&\quad\forall\,k\in[m],j\in[n],h\in[G], \label{desired:coeff1}\\
\mathbf{H}_{(k-1)pn+jp-1}^{h,\ell}&=&\sum\limits_{l=1}^{p}\mathbf{A}_{k,l}^{h,\ell}\mathbf{B}_{l,j}^{h,\ell}=\mathbf{C}_{k,j}^{h,\ell},&\quad\forall\,k\in[m],j\in[n],\ell\in[2:L], h\in[G], \label{desired:coeff2}
\end{IEEEeqnarray}
which are exactly the desired products $\mathbf{C}^{h,1}$ and $\mathbf{C}^{h,\ell}$ by \eqref{desred:product2}, respectively.
Define
\begin{IEEEeqnarray}{rCl}
\psi\triangleq(m-1)(np+X_{\mathbf{B}})+np,\quad 
\kappa\triangleq mpn,\label{def:kappa}
\end{IEEEeqnarray}
i.e., $\psi-1$ and $\kappa-1$ are the maximum indices of polynomial coefficients in $\{\mathbf{H}_r^{h,1}:r\in[0:\delta]\}$ and $\{\mathbf{H}_r^{h,\ell}:r\in[0:mpn+p-2]\}$ that can be directly used to recover $\mathbf{C}^{h,1}$ and $\mathbf{C}^{h,\ell}$ $(\ell\in[2:L])$ by \eqref{desired:coeff1} and \eqref{desired:coeff2}, respectively.
For convenience, we refer to $\{\mathbf{H}_r^{h,1}:r\in[0:\psi-1]\}$ and $\{\mathbf{H}_r^{h,\ell}:r\in[0:\kappa-1],\ell\in[2:L]\}$ with continuous indices as \emph{desired coefficients} for $h\in[G]$.

Then, the functions $\widetilde{\mathbf{A}}^h(\alpha)$ and $\widetilde{\mathbf{B}}^h(\alpha)$ are given by
\begin{IEEEeqnarray}{rCll}
\widetilde{\mathbf{A}}^{h}(\alpha)&=&\Delta^{h}(\alpha)\left( \frac{1}{(f_{h,1}-\alpha)^{\psi}}\mathbf{P}^{h,1}(\alpha)+\sum\limits_{\ell=2}^{L}\frac{1}{(f_{h,\ell}-\alpha)^{\kappa}}\mathbf{P}^{h,\ell}(\alpha) \right),& \quad\forall\,h\in[G], \label{security:multi-party}\\
\widetilde{\mathbf{B}}^{h}(\alpha)&=&
\frac{1}{(f_{h,1}-\alpha)^{\psi}}\mathbf{Q}^{h,1}(\alpha)+\sum\limits_{\ell=2}^{L}\frac{1}{(f_{h,\ell}-\alpha)^{\kappa}}\mathbf{Q}^{h,\ell}(\alpha),&\quad\forall\,h\in[G], \label{security:multi-party2}
\end{IEEEeqnarray}
where $\Delta^{h}(\alpha)\triangleq (f_{h,1}-\alpha)^{\psi}\cdot\prod_{\ell=2}^{L}(f_{h,\ell}-\alpha)^{\kappa}$, which is used for aligning the desired products $\{\mathbf{P}^{h,\ell}(\alpha)\mathbf{Q}^{h,\ell'}(\alpha):\ell,\ell'\in[L],\ell=\ell'\}$ along the Cauchy part and aligning the interference from the undesired cross products $\{\mathbf{P}^{h,\ell}(\alpha)\mathbf{Q}^{h,\ell'}(\alpha):\ell,\ell'\in[L],\ell\neq\ell'\}$ within dimensions corresponding to the Vandermonde part, as shown in the third term in \eqref{expanding form1234}.
Apparently, the two encoding functions satisfy the criterion C3$'$.

The key of our approach depends on the following results of expanding $\widetilde{\mathbf{A}}^{h}(\alpha)\cdot \widetilde{\mathbf{B}}^{h}(\alpha)$, which show the interference alignment rules and the interference are aligned along the Vandermonde terms of $\phi+1$ dimensions, where the parameter $\phi$ is given by
\begin{IEEEeqnarray}{c}
\phi\triangleq(L-1)mnp+np+X_{\mathbf{A}}+(m-1)X_{\mathbf{B}}-2.\label{definition:phi}
\end{IEEEeqnarray}
Thus, the criterion C2$'$ is satisfied. The detailed proof of Lemma \ref{lemma:prod} is relegated to Appendix.
%
%
\begin{Lemma}\label{lemma:prod} For any $h\in[G]$, $\widetilde{\mathbf{A}}^{h}(\alpha)\cdot \widetilde{\mathbf{B}}^{h}(\alpha)$  is given by the following form:
\begin{IEEEeqnarray}{rCl}
\widetilde{\mathbf{A}}^{h}(\alpha)\cdot \widetilde{\mathbf{B}}^{h}(\alpha)&=&\sum\limits_{r=0}^{\psi-1}\sum\limits_{s=r+1}^{\psi}\frac{c_{\psi-s}^{h,1}\cdot\mathbf{H}_r^{h,1}}{(f_{h,1}-\alpha)^{s-r}}+\sum\limits_{\ell=2}^{L}\sum\limits_{r=0}^{\kappa-1}\sum\limits_{s=r+1}^{\kappa}\frac{c_{\kappa-s}^{h,\ell}\cdot \mathbf{H}_r^{h,\ell}}{(f_{h,\ell}-\alpha)^{s-r}}+\sum\limits_{r=0}^{\phi}\mathbf{U}_{r}^{h}\alpha^{r},\label{expanding form1234}
\end{IEEEeqnarray}
where the coefficients $c_{0}^{h,1},c_1^{h,1},\ldots,c_{\psi-1}^{h,1}$ are determined by expanding $\prod_{\ell=2}^{L}\big(f_{h,\ell}-\alpha\big)^{\kappa}$ as a polynomial of $f_{h,1}-\alpha$, i.e.,
\begin{IEEEeqnarray}{c}
\prod_{\ell=2}^{L}\big(f_{h,\ell}-\alpha\big)^{\kappa}=\prod_{\ell=2}^{L}\big((f_{h,1}-\alpha)+(f_{h,\ell}-f_{h,1})\big)^{\kappa}=\sum\limits_{s=0}^{(L-1)\kappa}c_s^{h,1}(f_{h,1}-\alpha)^{s} , \label{coefficients:1}
\end{IEEEeqnarray}
and $c_s^{h,1}\triangleq0$ if $(L-1)\kappa<s<\psi$;
the coefficients $c_{0}^{h,\ell},c_1^{h,\ell},\ldots,c_{\kappa-1}^{h,\ell}$ ($\ell\in[2:L]$) are determined by expanding $\big(f_{h,1}-\alpha\big)^{\psi}\cdot\prod_{k\in[2:L]\backslash\{\ell\}}\big(f_{h,k}-\alpha\big)^{\kappa}$ as the polynomial of $f_{h,\ell}-\alpha$, i.e.,
\begin{IEEEeqnarray}{rCl}
\big((f_{h,\ell}-\alpha)+(f_{h,1}-f_{h,\ell})\big)^{\psi}\cdot\prod_{k\in[2:L]\backslash\{\ell\}}\big((f_{h,\ell}-\alpha)+(f_{h,k}-f_{h,\ell})\big)^{\kappa}
&=&\sum\limits_{s=0}^{(L-2)\kappa+\psi}c_s^{h,\ell}(f_{h,\ell}-\alpha)^{s};\IEEEeqnarraynumspace \label{coefficients:2}
\end{IEEEeqnarray}
and the matrices $\{\mathbf{U}_r^h\}_{r=0}^{\phi}$ with sizes of $\frac{\lambda}{m}\times\frac{\vartheta}{n}$ are determined by various linear combinations of the products of sub-matrices in group $h$, whose exact forms are unimportant.


\end{Lemma}

\textbf{Sharing:} Let $\alpha_1,\ldots,\alpha_N$ be $N$ distinct elements from $\mathbb{F}_q\backslash\{f_{h,\ell}:h\in[G],\ell\in[L]\}$. Source $1$ and Source $2$ share the evaluations of $\widetilde{\mathbf{A}}^{1}(\alpha),\ldots,\widetilde{\mathbf{A}}^{G}(\alpha)$ and $\widetilde{\mathbf{B}}^{1}(\alpha),\ldots,\widetilde{\mathbf{B}}^{G}(\alpha)$ at $\alpha=\alpha_i$ with server $i$, respectively, i.e.,
\begin{IEEEeqnarray}{c}
\widetilde{\mathbf{A}}_{i}=(\widetilde{\mathbf{A}}^{1}(\alpha_{i}),\ldots,\widetilde{\mathbf{A}}^{G}(\alpha_{i})),\quad
\widetilde{\mathbf{B}}_{i}=(\widetilde{\mathbf{B}}^{1}(\alpha_{i}),\ldots,\widetilde{\mathbf{B}}^{G}(\alpha_{i})).\label{sharing:B}
\end{IEEEeqnarray}

\textbf{Computation:} Define $\Gamma$ and $\Lambda$ as the sets that consist of indices of the coefficients in \eqref{desired:coeff1} and \eqref{desired:coeff2}, respectively, i.e., $\Gamma\triangleq\{(k-1)(pn+X_{\mathbf{B}})+jp-1:k\in[m],j\in[n]\}$ and
$\Lambda\triangleq\{(k-1)pn+jp-1:k\in[m],j\in[n]\}$.
To ensure the security of $\mathbf{A},\mathbf{B}$ for the user beyond its desired products, let the random variables shared by all the servers be
\begin{IEEEeqnarray}{rCl}\label{random:sharing2}
\mathcal{S}&=&\{\mathbf{Z}_{r}^{h,1}:r\in[0:\psi-1]\backslash\Gamma,h\in[G]\} \notag \\
&&\quad\quad\quad\quad\quad\quad\quad\cup\{\mathbf{Z}_{r}^{h,\ell}:r\in[0:\kappa-1]\backslash\Lambda,\ell\in[2:L],h\in[G]\}\cup\{\mathbf{Z}_r:r\in[0:\phi]\},\IEEEeqnarraynumspace
\end{IEEEeqnarray}
chosen independently and uniformly from all matrices over $\mathbb{F}_q^{\frac{\lambda}{m}\times\frac{\vartheta}{n}}$.
In addition, for the indices in $\Gamma$ and $\Lambda$, define
\begin{IEEEeqnarray}{rCl}
\mathbf{Z}_{r}^{h,1}&\triangleq&[\mathbf{0}]_{\frac{\lambda}{m}\times\frac{\vartheta}{n}},\quad\forall\,r\in\Gamma,\ell=1,h\in[G], \label{set:zero1}\\
\mathbf{Z}_{r}^{h,\ell}&\triangleq&[\mathbf{0}]_{\frac{\lambda}{m}\times\frac{\vartheta}{n}},\quad\forall\, r\in\Lambda,\ell\in[2:L],h\in[G].\label{set:zero2}
\end{IEEEeqnarray}
According to the criterion C4$'$, define a function of $\alpha$ similar structure to \eqref{expanding form1234}:
\begin{IEEEeqnarray}{rCl}
\mathbf{S}(\alpha)&\triangleq&\sum\limits_{h=1}^{G}\sum\limits_{r=0}^{\psi-1}\sum\limits_{s=r+1}^{\psi}\frac{c_{\psi-s}^{h,1}}{(f_{h,1}-\alpha)^{s-r}}\mathbf{Z}_r^{h,1}
+\sum\limits_{h=1}^{G}\sum\limits_{\ell=2}^{L}\sum\limits_{r=0}^{\kappa-1}\sum\limits_{s=r+1}^{\kappa}\frac{c_{\kappa-s}^{h,\ell}}{(f_{h,\ell}-\alpha)^{s-r}}\mathbf{Z}_r^{h,\ell}+\sum\limits_{r=0}^{\phi}\mathbf{Z}_{r}\alpha^{r}.\IEEEeqnarraynumspace\label{def:S:alp}
\end{IEEEeqnarray}

Then, server $i$ computes a response by taking the inner product of the received sub-matrix vectors $\widetilde{\mathbf{A}}_{i}$ and $\widetilde{\mathbf{B}}_{i}$, and adding on an evaluation of $\mathbf{S}(\alpha)$ at $\alpha=\alpha_i$, i.e.,
\begin{IEEEeqnarray}{rCl}
\mathbf{Y}_{i}&=&\sum\limits_{h=1}^{G}\widetilde{\mathbf{A}}^{h}(\alpha_i)\cdot\widetilde{\mathbf{B}}^{h}(\alpha_i)+\mathbf{S}(\alpha_i) \label{answer cost2}\\
&=&\sum\limits_{h=1}^{G}\sum\limits_{r=0}^{\psi-1}\sum\limits_{s=r+1}^{\psi}\frac{c_{\psi-s}^{h,1}(\mathbf{H}_r^{h,1}+\mathbf{Z}_r^{h,1})}{(f_{h,1}-\alpha_i)^{s-r}}\notag \\
&&\quad\quad\quad\quad\quad\quad\quad\quad+\sum\limits_{h=1}^{G}\sum\limits_{\ell=2}^{L}\sum\limits_{r=0}^{\kappa-1}\sum\limits_{s=r+1}^{\kappa}\frac{c_{\kappa-s}^{h,\ell}(\mathbf{H}_r^{h,\ell}+\mathbf{Z}_r^{h,\ell})}{(f_{h,\ell}-\alpha_i)^{s-r}}+
\sum\limits_{r=0}^{\phi}(\mathbf{U}_r+\mathbf{Z}_r)\alpha_i^r,  \IEEEeqnarraynumspace\label{servers:responses}
\end{IEEEeqnarray}
where $\mathbf{U}_r\triangleq\sum_{h=1}^{G}\mathbf{U}_r^{h}$  for any $r\in[0:\phi]$ and the evaluation follows from Lemma \ref{lemma:prod} and \eqref{def:S:alp}. By \eqref{set:zero1} and \eqref{set:zero2}, the responses reserve completely the terms $\{\mathbf{H}_r^{h,1}:r\in\Gamma,h\in[G]\}$ and $\{\mathbf{H}_r^{h,\ell}:r\in\Lambda,\ell\in[2:L],h\in [G]\}$, and uses random matrices to scramble all the residual terms.

Moreover, in the server responses \eqref{servers:responses}, the desired coefficients $\mathbf{H}_r^{h,1}$ $(r\in[0:\psi-1])$ and $\mathbf{H}_r^{h,\ell}$ $(r\in[0:\kappa-1],\ell\in[2:L])$ appear along the Cauchy terms $\sum_{s=r+1}^{\psi}\frac{c_{\psi-s}^{h,1}}{(f_{h,1}-\alpha_i)^{s-r}}$ and $\sum_{s=r+1}^{\kappa}\frac{c_{\kappa-s}^{h,\ell}}{(f_{h,\ell}-\alpha_i)^{s-r}}$ for each group $h$, respectively, and everything else (interference) are always distributed over the Vandermonde terms $\{\alpha^{r}_i:r\in[0:\phi]\}$ of $\phi+1$ dimensions and the number of interference dimensions has not increased at all while we sum the matrix products over the $G$ groups, compared to \eqref{expanding form1234}.
However, the upload and server computation complexity are scaled with $G$. Thus, such grouping method is instrumental in focusing on the general tradeoff with respect to recovery threshold, system cost and complexity.

\textbf{Reconstruction: } The responses from any $K=G\psi+G(L-1)\kappa+\phi+1$
servers, whose indices are denoted as $i_1,i_2,\ldots,i_K$, 
can be written in matrix form as follow. 
\begin{figure}[htp]
\begin{IEEEeqnarray}{c}
\left[
\begin{array}{@{}c@{}}
\mathbf{Y}_{i_1} \\
\mathbf{Y}_{i_2} \\
\vdots \\
\mathbf{Y}_{i_K}
\end{array}
\right]=\big((
\underbrace{
\left[
\begin{array}{ccc;{2pt/2pt}c;{2pt/2pt}ccc;{2pt/2pt}cccc}
\frac{1}{(f_{1,1}-\alpha_{i_1})^{\psi}} & \ldots & \frac{1}{f_{1,1}-\alpha_{i_1}} & \ldots & \frac{1}{(f_{G,L}-\alpha_{i_1})^{\kappa}} & \ldots & \frac{1}{f_{G,L}-\alpha_{i_1}} & 1 & \alpha_{i_1} & \ldots &\alpha_{i_1}^{\phi} \\
\frac{1}{(f_{1,1}-\alpha_{i_2})^{\psi}} & \ldots & \frac{1}{f_{1,1}-\alpha_{i_2}} & \ldots & \frac{1}{(f_{G,L}-\alpha_{i_2})^{\kappa}} & \ldots & \frac{1}{f_{G,L}-\alpha_{i_2}} & 1 & \alpha_{i_2} & \ldots & \alpha_{i_2}^{\phi} \\
\vdots & \vdots & \vdots & \vdots & \vdots & \vdots & \vdots & \vdots & \vdots & \vdots & \vdots \\
\frac{1}{(f_{1,1}-\alpha_{i_K})^{\psi}} & \ldots & \frac{1}{f_{1,1}-\alpha_{i_K}} & \ldots & \frac{1}{(f_{G,L}-\alpha_{i_K})^{\kappa}} & \ldots & \frac{1}{f_{G,L}-\alpha_{i_K}} & 1 & \alpha_{i_K} & \ldots & \alpha_{i_K}^{\phi}
\end{array}
\right]
}_{\mathbf{V}_1} \notag\\
\underbrace{
\left[
\begin{array}{c;{2pt/2pt}c;{2pt/2pt}c;{2pt/2pt}c}
\textbf{T}(c_{0}^{1,1},\ldots,c_{\psi-1}^{1,1})&&& \\ \hdashline[2pt/2pt]
&\ddots & &\\\hdashline[2pt/2pt]
& & \textbf{T}(c_{0}^{G,L},\ldots,c_{\kappa-1}^{G,L})& \\\hdashline[2pt/2pt]
& & & \mathbf{I}_{\phi+1} \\
\end{array}
\right]}_{\mathbf{V}_2})
\otimes\mathbf{I}_{\lambda/m}\big)
\left[
\begin{array}{@{}c@{}}
\mathbf{H}_0^{1,1}+\mathbf{Z}_0^{1,1} \\
\vdots \\
\mathbf{H}_{\psi-1}^{1,1}+\mathbf{Z}_{\psi-1}^{1,1} \\\hdashline[2pt/2pt]
\vdots \\\hdashline[2pt/2pt]
\mathbf{H}_0^{G,L}+\mathbf{Z}_0^{G,L} \\
\vdots \\
\mathbf{H}_{\kappa-1}^{G,L}+\mathbf{Z}_{\kappa-1}^{G,L} \\\hdashline[2pt/2pt]
\mathbf{U}_0+\mathbf{Z}_0 \\
\mathbf{U}_1+\mathbf{Z}_1 \\
\vdots \\
\mathbf{U}_{\phi}+\mathbf{Z}_{\phi} \\
\end{array}
\right]. \label{answers:multi party2}
\end{IEEEeqnarray}
\end{figure}

Recall that $f_{1,1},f_{1,2},\ldots,f_{G,L},\alpha_1,\ldots,\alpha_N$ are distinct elements from $\mathbb{F}_q$.  From \eqref{coefficients:1} and \eqref{coefficients:2},
\begin{IEEEeqnarray*}{rCl}
c_{0}^{h,1}&=&\prod_{\ell=2}^{L}(f_{h,\ell}-f_{h,1})^{\kappa},\quad\forall\,\ell=1,h\in[G], \\
c_{0}^{h,\ell}&=&(f_{h,1}-f_{h,\ell})^{\psi}\cdot\prod_{k\in[2:L]\backslash\{\ell\}}(f_{h,k}-f_{h,\ell})^{\kappa},\quad\forall\,\ell\in[2:L],h\in[G],
\end{IEEEeqnarray*}
and thus $c_{0}^{1,1},\ldots,c_0^{G,L}$ are all non-zero elements. Following by Lemma \ref{C-V matrix pole} and the fact the Kronecker product of non-singular matrices is non-singular, the Cauchy-Vandermonde matrix $\mathbf{V}_1\mathbf{V}_2\otimes\mathbf{I}_{\lambda/m}$ is invertible, which ensures that the desired coefficients are separated from each other and interference, i.e., the criterion C1$'$ is satisfied.

By inverting $\mathbf{V}_1$ and $\mathbf{V}_2$, from the responses of any $K$ servers, the user can decode
\begin{IEEEeqnarray*}{c}
\Big\{\mathbf{H}_r^{h,1}+\mathbf{Z}_{r}^{h,1}:r\in[0:\psi-1],h\in[G]\Big\}\cup\Big\{\mathbf{H}_r^{h,\ell}+\mathbf{Z}_{r}^{h,\ell}:r\in[0:\kappa-1],\ell\in[2:L],h\in[G]\Big\}.
\end{IEEEeqnarray*}
By \eqref{set:zero1} and \eqref{set:zero2}, we further obtain $\mathbf{H}_{r}^{h,1},r\in\Gamma,h\in[G]$ and $\mathbf{H}_{r}^{h,\ell},r\in\Lambda,\ell\in[2:L],h\in[G]$.
Thus, the desired product $\mathbf{A}\divideontimes\mathbf{B}=(\mathbf{A}^{1,1}\mathbf{B}^{1,1},\ldots,\mathbf{A}^{G,L}\mathbf{B}^{G,L})$ are recovered by \eqref{desired:coeff1} and \eqref{desired:coeff2}.

Consequently, by \eqref{def:kappa} and \eqref{definition:phi}, the computation strategy achieves the recovery threshold
\begin{IEEEeqnarray}{rCl}
K&=&G\psi+G(L-1)\kappa+\phi+1 =(LG+L-1)mpn+np+X_{\mathbf{A}}+(G+1)(m-1)X_{\mathbf{B}}-1.\notag
\end{IEEEeqnarray}

Similar to \eqref{encoding:function:1} and \eqref{encoding:function:2}, we can exchange the two encoding functions of $\mathbf{A}$ and $\mathbf{B}$, i.e., for any $h\in[G]$, let the encoding functions of matrices $\mathbf{A}^{h,\ell}$ and $\mathbf{B}^{h,\ell}$ be
\begin{IEEEeqnarray*}{rCl}
\mathbf{P}^{h,1}(\alpha)&=&\sum\limits_{k=1}^{m}\left(\sum\limits_{l=1}^{p}\mathbf{A}_{k,l}^{h,1}(f_{h,1}-\alpha)^{l-1}\right)(f_{h,1}-\alpha)^{(k-1)p}+\sum\limits_{x=1}^{X_{\mathbf{A}}}\mathbf{Z}_{x}^{\mathbf{A},h}(f_{h,1}-\alpha)^{mp+(x-1)},\\
\mathbf{Q}^{h,1}(\alpha)&=&\sum\limits_{j=1}^{n}\left(\sum\limits_{l=1}^{p}\mathbf{B}_{l,j}^{h,1}(f_{h,1}-\alpha)^{p-l}\right)(f_{h,1}-\alpha)^{(j-1)(mp+X_{\mathbf{A}})}\notag\\
&&\quad\quad\quad\quad\quad\quad\quad\quad\quad\quad\quad\quad+\sum\limits_{x=1}^{X_{\mathbf{B}}}\mathbf{Z}_{x}^{\mathbf{B},h}(f_{h,1}-\alpha)^{(n-1)(mp+X_{\mathbf{A}})+mp+(x-1)},\\
\mathbf{P}^{h,\ell}(\alpha)&=&\sum\limits_{k=1}^{m}\left(\sum\limits_{l=1}^{p}\mathbf{A}_{k,l}^{h,\ell}(f_{h,\ell}-\alpha)^{l-1}\right){(f_{h,\ell}-\alpha)^{(k-1)p}}, \quad\forall\, \ell\in[2:L]. \\
\mathbf{Q}^{h,\ell}(\alpha)&=&\sum\limits_{j=1}^{n}\left(\sum\limits_{l=1}^{p}\mathbf{B}_{l,j}^{h,\ell}(f_{h,\ell}-\alpha)^{p-l}\right){(f_{h,\ell}-\alpha)^{(j-1)mp}},\quad\forall\,\ell\in[2:L].
\end{IEEEeqnarray*}
Then, following the similar arguments to the above computation strategy for SMBMM, it is straightforward to obtain another computation strategy that achieves the recovery threshold
\begin{IEEEeqnarray}{c}
K=(LG+L-1)mpn+mp+X_{\mathbf{B}}+(G+1)(n-1)X_{\mathbf{A}}-1.\notag
\end{IEEEeqnarray}
\begin{Theorem}\label{main result:2}
For an SMBMM problem
with parameters $m,p,n,M,X_{\mathbf{A}},X_{\mathbf{B}},G,L\in\mathbb{Z}^{+}$ such that $m>1,n>1,L>1$ and $M=GL$, the proposed computation strategy achieves
\begin{IEEEeqnarray*}{rCl}
\text{Recovery Threshold:}\quad& K=\min\{K',K''\}, \\
\text{Upload Cost of Sources:}\quad& (U_{\mathbf{A}},U_{\mathbf{B}})=\left(\frac{N}{Lmp},\frac{N}{Lnp}\right),\\
\text{Amount of Common Randomness:}\quad&  \rho=\frac{K}{GLmn}-1,\\
\text{Download Cost:}\quad&  D=\frac{K}{GLmn}, \\
\text{Encoding Complexity:}\quad& (\mathcal{C}_{\mathbf{A}},\mathcal{C}_{\mathbf{B}})=\left(\widetilde{\mathcal{O}}\left(\frac{\lambda\xi N(\log N)^{2}}{Lmp}\right),\widetilde{\mathcal{O}}\left(\frac{\xi\vartheta N(\log N)^{2}}{Lnp}\right)\right),   \\
\text{Server Computation Complexity:}\quad& \mathcal{C}_s=\mathcal{O}\left(\frac{\lambda\xi\vartheta}{Lmpn}\right), \\
\text{Decoding Complexity:}\quad& \mathcal{C}_d=\widetilde{\mathcal{O}}\left(\frac{\lambda\vartheta K(\log K)^{2}}{LGmn}\right),
\end{IEEEeqnarray*}
where $K'=(LG+L-1)mpn+np+X_{\mathbf{A}}+(G+1)(m-1)X_{\mathbf{B}}-1$ and $K''=(LG+L-1)mpn+mp+X_{\mathbf{B}}+(G+1)(n-1)X_{\mathbf{A}}-1$.
\end{Theorem}
\begin{IEEEproof}
Obviously, the recovery threshold $\min\{K',K''\}$ can be achieved by the above computation strategies for SMBMM. Its securities, system cost and complexities are proved in Section \ref{proof:strategy:2}.
\end{IEEEproof}

\begin{Remark}\label{footenote 1}
When $L=1$, following similar arguments to the case of $L\geq 2$, the computation strategy with recovery threshold $\min\{Gmpn+np+X_{\mathbf{A}}+(Gm-G+1)X_{\mathbf{B}}-1,Gmpn+mp+X_{\mathbf{B}}+(Gn-G+1)X_{\mathbf{A}}-1\}$ can be obtained straightly. Here, we omit them due to space limit.
\end{Remark}

\begin{Remark}\label{diff}
The SMBMM problem was first introduced in \cite{Chen and Jafar} under the special case $X=X_{\mathbf{A}}=X_{\mathbf{B}}$ and a computation strategy \cite{Chen and Jafar} with recovery threshold $(LG+L)mpn+2X-1$ is presented.
Apparently, the recovery threshold of our strategy outperforms \cite{Chen and Jafar} when $X\leq\frac{\max\{np,mp\}}{G+1}$, see Section \ref{section:comparison} for detailed comparisons. 
The main difference between the strategy in \cite{Chen and Jafar} and ours is that the encoding functions in \cite{Chen and Jafar} \emph{separately} use Cauchy coefficients to encode the partitioning sub-matrices of data matrices $\mathbf{A}^{h,\ell},\mathbf{B}^{h,\ell}$ and use Vandermonde coefficients to encode random noise matrices, specifically, we design appropriate Cauchy coefficients to \emph{jointly} encode the noise matrices and the sub-matrices in $\mathbf{A}^{h,\ell},\mathbf{B}^{h,\ell}$ \eqref{security:multi-party}-\eqref{security:multi-party2}, which creates more interference alignment opportunities (see Table \ref{tab:dim align}) when $X\leq\frac{\max\{np,mp\}}{G+1}$.

\begin{table*}[htbp]\centering
\caption{{Comparison for dimensions of interference alignment of SMBMM strategies}}\label{tab:dim align}
  \begin{tabular}{|c|c|c|}
  \hline
  & \multirow{2}*{\minitab[c]{Dimensions occupied \\ by desired products}}  & \multirow{2}*{\minitab[c]{Dimensions occupied \\ by interference}} \\
  &  &  \\ \hline
    Our Degraded Strategy & $LGmpn$ & $K_{int}$ \\ \hline
  Chen \emph{et al.} Strategy \cite{Chen and Jafar} & $LGmpn$ & $Lmpn+2X-1$ \\ \hline
  \end{tabular}
\begin{tablenotes}
       \footnotesize
       \item[]\quad Here, $K_{int}=\min\{(L-1)mpn+np+(Gm-G+m)X-1, (L-1)mpn+mp+(Gn-G+n)X-1\}$.
\end{tablenotes}
\end{table*}
\end{Remark}

\subsection{Securities, System Cost and Complexities for SMBMM}\label{proof:strategy:2}
Since the two computation strategies with recovery thresholds $K'$ and $K''$ are similar, it is enough to focus on the former.

\subsubsection{Securities}
Similar to the computation Strategy of SSMM in Section \ref{construction_SDMM}, it is easy to prove that the sharing phases \eqref{sharing:B} satisfy the security constraints \eqref{sec const}, \eqref{sec const2}.
We know from \eqref{def:S:alp} and \eqref{servers:responses} that the noise polynomial $\mathbf{S}(\alpha)$ is constructed in the same structure as $\sum_{h=1}^{G}\widetilde{\mathbf{A}}^{h}(\alpha_i)\cdot\widetilde{\mathbf{B}}^{h}(\alpha_i)$.
Thus, no information about the data matrices $\mathbf{A},\mathbf{B}$ is leaked to the user beyond the desired products $\mathbf{A}\divideontimes\mathbf{B}$, i.e., the security for the user \eqref{security for user} is satisfied.

\subsubsection{System Cost}
By \eqref{sharing:B}, Source $1$ and Source $2$ share $G$ encoding matrices with sizes of $\frac{\lambda\xi}{mp}$ and $\frac{\xi\vartheta}{np}$ to each server, respectively. Thus, the normalized upload cost are
$U_{\mathbf{A}}=\frac{NG\frac{\lambda\xi}{mp}}{GL\lambda\xi}=\frac{N}{Lmp}$ and $U_{\mathbf{B}}=\frac{NG\frac{\xi\vartheta}{np}}{GL\xi\vartheta}=\frac{N}{Lnp}$.
By \eqref{answer cost2}, the answer of each server is a matrix of size $\frac{\lambda}{m}\times\frac{\vartheta}{n}$. Thus, from the responses of any $K'$ servers, the normalized download cost is
$D=\frac{K'\frac{\lambda}{m}\times\frac{\vartheta}{n}}{GL\lambda\vartheta}=\frac{K'}{GLmn}$.
From \eqref{random:sharing2}, the amount of shared common variables is
\begin{IEEEeqnarray*}{c}
\rho=\frac{(GL+L-1)mpn-GLmn+np+X_{\mathbf{A}}+(G+1)(m-1)X_{\mathbf{B}}-1}{GLmn}=\frac{K'}{GLmn}-1.
\end{IEEEeqnarray*}

\subsubsection{Complexity Analysis}For the server computation complexity \eqref{answer cost2}, each server multiplies $G$ pairs of encoding sub-matrices with sizes of $\frac{\lambda}{m}\times\frac{\xi}{p}$ and $\frac{\xi}{p}\times\frac{\vartheta}{n}$ and evaluates $\mathbf{S}(\alpha)$ at one point, and then returns the sum of these $G$ products and the evaluation. By using straightforward matrix multiplication, the complexity of the $G$ pairs of matrices multiplication is at most $\mathcal{O}(\frac{\lambda\xi\vartheta}{Lmpn})$.
Evaluating of $\mathbf{S}(\alpha)$ at one point can be viewed as computing a linear combination of $K'$ matrices of size $\frac{\lambda}{m}\times\frac{\vartheta}{n}$ and achieves the complexity $\mathcal{O}(\frac{\lambda\vartheta K'}{GLmn})$, since the coefficients $\big\{\sum_{s=r+1}^{\psi}\frac{c_{\psi-s}^{h,1}}{(f_{h,1}-\alpha_i)^{s-r}}:r\in[0:\psi-1],h\in[G]\big\}$ and $\big\{\sum_{s=r+1}^{\kappa}\frac{c_{\kappa-s}^{h,\ell}}{(f_{h,\ell}-\alpha_i)^{s-r}}:r\in[0:\kappa],h\in[G]\big\}$ in $\mathbf{S}(\alpha_i)$ are independent of the computation strategy and thus can be computed at server $i$ a priori during off-peak hours to reduce the latency of server computation.
And the complexity of summation over the $G$ products and evaluation is $\mathcal{O}(\frac{\lambda\vartheta}{Lmn})$. Thus, the computation complexity at each server is at most $\mathcal{O}(\frac{\lambda\xi\vartheta}{Lmpn})+{\mathcal{O}}(\frac{\lambda\vartheta K'}{GLmn})+\mathcal{O}(\frac{\lambda\vartheta}{Lmn})$, which is dominated by  $\mathcal{O}(\frac{\lambda\xi\vartheta}{Lmpn})$ because the data matrices (i.e., $\lambda,\xi,\vartheta$) are sufficiently large in practical scenarios.
The encoding and decoding complexities follow the same arguments as G-CSA codes \cite{Chen and Jafar,Jia and Jafar}, which are at most $(\mathcal{C}_{\mathbf{A}},\mathcal{C}_{\mathbf{B}})=(\widetilde{\mathcal{O}}(\frac{\lambda\xi N(\log N)^2}{Lmp}),\widetilde{\mathcal{O}}(\frac{\xi\vartheta N(\log N)^2}{Lnp}))$ and $\widetilde{\mathcal{O}}(\frac{\lambda\vartheta K'(\log K')^2}{LGmn})$, respectively.

\section{Performance Comparisons}\label{section:comparison}

Both traditional SSMM and SMBMM 
mainly focus on the special case that the matrices $\mathbf{A}$ and $\mathbf{B}$ have the same security level (i.e., $X=X_{\mathbf{A}}=X_{\mathbf{B}}$).
For SSMM, the state of the art are reflected in Secure Generalized PolyDot (S-GPD) codes \cite{GSPolyDot code},
GASP codes \cite{Rouayheb secure code}, A3S codes \cite{Kakar secure code}, USCSA codes \cite{Kakar and Khristoforov}, FFT-based Polynomial (FFT-P) codes \cite{Gunduz20}, Polynomial Sharing (PS) codes \cite{EP SMC} and Extended Entangled Polynomial (E-EP) codes \cite{EP code}.
Depending on the partitioning manners of the data matrices, these works are divided into two classes of row-by-column partition \cite{Rouayheb secure code,Kakar secure code,Kakar and Khristoforov} (where the partitioning parameter $p$ is set to be $1$ and $m,n$ are arbitrary) and arbitrary partition \cite{GSPolyDot code,Gunduz20,EP SMC,EP code} (where all the partitioning parameters $m,p,n$ are arbitrary).
%
For ease of comparison, we list  the performance of the aforementioned codes and our degraded strategies in Tables \ref{tab:comparative123} and \ref{tab:comparative456}.

\begin{table*}[htbp]
\centering
\small
\caption{{Comparison for SSMM strategies with row-by-column partition of data matrices}}\label{tab:comparative123}
\resizebox{145mm}{34mm}{
  \begin{tabular}{|c|c|c|c|}
  \hline
  & \multirow{2}*{\minitab[c]{Recovery Threshold \\ $K$}}  & \multirow{2}*{\minitab[c]{Upload Cost \\ $(U_{\mathbf{A}},U_{\mathbf{B}})$}}  & \multirow{2}*{\minitab[c]{Download Cost \\ $D$}} \\
  &  &  &  \\ \hline
  GASP Codes \cite{Rouayheb secure code} & $K_{GP}$ & $\left(\frac{N}{m},\frac{N}{n}\right)$ & $\frac{K_{GP}}{mn}$ \\ \hline
  A3S Codes \cite{Kakar secure code} & $K_{AS}=(n+1)(m+X)-1$ &  $\left(\frac{N}{m},\frac{N}{n}\right)$ & $\frac{K_{AS}}{mn}$ \\ \hline
  USCSA(0) Codes \cite{Kakar and Khristoforov} & $K_{U0}=mn+m+2X-1$ & $\left(\frac{Nn}{m},N\right)$ & $\frac{K_{U0}}{mn}$ \\ \hline
  USCSA(1) Codes \cite{Kakar and Khristoforov} & $K_{U1}=mn+n+2X-1$ & $\left(N,\frac{Nm}{n}\right)$ & $\frac{K_{U1}}{mn}$ \\ \hline
  Degraded Strategy-1  & $K_{D1}$ & $\left(\frac{N}{m},\frac{N}{n}\right)$ & $\frac{K_{D1}}{mn}$ \\ \hline\hline
  & Encoding Complexity & Server Computation & Decoding Complexity \\
  & $(\mathcal{C}_{\mathbf{A}},\mathcal{C}_{\mathbf{B}})$ & Complexity $\mathcal{C}_s$ & $\mathcal{C}_d$ \\ \hline
  GASP Codes \cite{Rouayheb secure code} & $\left(\widetilde{\mathcal{O}}\left(\frac{\lambda\xi N(\log N)^{2}}{m}\right),\widetilde{\mathcal{O}}\left(\frac{\xi\vartheta N(\log N)^2}{n}\right)\right)$ & $\mathcal{O}\left(\frac{\lambda\xi\vartheta}{mn}\right)$ & ${\mathcal{O}}\left(\frac{\lambda\vartheta K_{GP}^2}{mn}\right)$ \\ \hline
  A3S Codes \cite{Kakar secure code} & $\left(\widetilde{\mathcal{O}}\left(\frac{\lambda\xi N(\log N)^{2}}{m}\right),\widetilde{\mathcal{O}}\left(\frac{\xi\vartheta N(\log N)^2}{n}\right)\right)$ & $\mathcal{O}\left(\frac{\lambda\xi\vartheta}{mn}\right)$ & $\widetilde{\mathcal{O}}\left(\frac{\lambda\vartheta K_{AS}(\log K_{AS})^2}{mn}\right)$ \\ \hline
  USCSA(0) Codes \cite{Kakar and Khristoforov} & $\left(\widetilde{\mathcal{O}}\left(\frac{n\lambda\xi N(\log N)^{2}}{m}\right),\widetilde{\mathcal{O}}\left(\xi\vartheta N(\log N)^2\right)\right)$ & $\mathcal{O}\left(\frac{\lambda\xi\vartheta}{m}\right)$ & $\widetilde{\mathcal{O}}\left(\frac{\lambda\vartheta K_{U0}(\log K_{U0})^2}{mn}\right)$ \\ \hline
  USCSA(1) Codes \cite{Kakar and Khristoforov} & $\left(\widetilde{\mathcal{O}}\left(\lambda\xi N(\log N)^{2}\right),\widetilde{\mathcal{O}}\left(\frac{m\xi\vartheta N(\log N)^2}{n}\right)\right)$ & $\mathcal{O}\left(\frac{\lambda\xi\vartheta}{n}\right)$ & $\widetilde{\mathcal{O}}\left(\frac{\lambda\vartheta K_{U1}(\log K_{U1})^2}{mn}\right)$ \\ \hline
  Degraded Strategy-1  & $\left(\widetilde{\mathcal{O}}\left(\frac{\lambda\xi N(\log N)^{2}}{m}\right),\widetilde{\mathcal{O}}\left(\frac{\xi\vartheta N(\log N)^2}{n}\right)\right)$ & $\mathcal{O}\left(\frac{\lambda\xi\vartheta}{mn}\right)$ & $\widetilde{\mathcal{O}}\left(\frac{\lambda\vartheta K_{D1}(\log K_{D1})^2}{mn}\right)$ \\ \hline
  \end{tabular}}
\begin{tablenotes}
       \footnotesize
       \item[]\quad\quad\quad\,\, Here, $K_{GP}=
      \left\{\begin{array}{@{}ll}
                                mn+m+n,&\mathrm{if}~1=X<n\leq m \\
                                mn+m+n+X^2+X-3,  &\mathrm{if}~2\leq X<n\leq m \\
                                (m+X)(n+1)-1,  &\mathrm{if}~n\leq m\leq x\\
                                \end{array}
      \right.$ for $n\leq m$ and  $K_{GP}$ is given
      \item[]\quad\quad\quad\,\, by interchanging $m$ and $n$ for $m<n$. Moreover, $K_{D1}=\min\{(m+1)(n+X),(n+1)(m+X)\}-1$.
     \end{tablenotes}
\end{table*}

For the case of row-by-column partition, we have  the following observations from Table \ref{tab:comparative123}:
\begin{enumerate}
\item In terms of recovery threshold, GASP codes outperform our degraded strategy if $2\leq X<\min\{m,n\}$ or $X\geq\max\{m,n\}$, and are the same as ours otherwise.
However, the decoding operation in GASP codes requires the inversion of a generalized Vandermonde matrix as opposed to the standard Vandermonde matrix inversion in our strategy. This results in
a higher decoding complexity and requires a sufficiently large finite field to ensure decodability and security \cite{Rouayheb secure code}. 
\item When $n>m$, our degraded strategy-1 achieves better recovery threshold, download cost and decoding complexity than A3S codes, while the other performance measures remain the same.
\item USCSA(0) codes and USCSA(1) codes outperform our strategy in terms of recovery threshold, download cost and decoding complexity, but it expanses higher upload cost, encoding complexity and server computation complexity.
\end{enumerate}

\begin{table*}[htbp]
\centering
\caption{{Comparison for SSMM strategies with arbitrary partition of data matrices}}\label{tab:comparative456}
\resizebox{140mm}{40mm}{
  \begin{tabular}{|c|c|c|c|}
  \hline
  & Recovery Threshold & Upload Cost & Download Cost \\
  & $K$ & $(U_{\mathbf{A}},U_{\mathbf{B}})$ & $D$ \\ \hline
  S-GPD Codes \cite{GSPolyDot code} & $K_{SD}$ & $\left(\frac{N}{mp},\frac{N}{np}\right)$ & $\frac{K_{SD}}{mn}$ \\ \hline
  PS codes \cite{EP SMC}  & $K_{PS}=2mpn+2X-1$ &  $\left(\frac{N}{mp},\frac{N}{np}\right)$ & $\frac{K_{PS}}{mn}$ \\ \hline
  DFT-P codes \cite{Gunduz20}  & $K_{DP}=mpn+2mnX$ & $\left(\frac{N}{mp},\frac{N}{np}\right)$ & $\frac{K_{DP}}{mn}$ \\ \hline
  E-EP codes \cite{EP code} & $K_{EP}=2R(m,p,n)+2X-1$ & $\left(\frac{N}{mp},\frac{N}{np}\right)$ & $\frac{K_{EP}}{mn}$ \\ \hline
  Degraded Strategy-2  & $K_{D2}$ & $\left(\frac{N}{mp},\frac{N}{np}\right)$ & $\frac{K_{D2}}{mn}$ \\ \hline\hline
  & Encoding Complexity & Server Computation & Decoding Complexity \\
  & $(\mathcal{C}_{\mathbf{A}},\mathcal{C}_{\mathbf{B}})$ & Complexity $\mathcal{C}_s$ & $\mathcal{C}_d$ \\ \hline
  S-GPD Codes \cite{GSPolyDot code} & $\left(\widetilde{\mathcal{O}}\left(\frac{\lambda\xi N(\log N)^{2}}{mp}\right),\widetilde{\mathcal{O}}\left(\frac{\xi\vartheta N(\log N)^2}{np}\right)\right)$ & $\mathcal{O}\left(\frac{\lambda\xi\vartheta}{mpn}\right)$ & $\widetilde{\mathcal{O}}\left(\frac{\lambda\vartheta K_{SD}(\log K_{SD})^2}{mn}\right)$ \\ \hline
  PS codes \cite{EP SMC}  & $\left(\widetilde{\mathcal{O}}\left(\frac{\lambda\xi N(\log N)^{2}}{mp}\right),\widetilde{\mathcal{O}}\left(\frac{\xi\vartheta N(\log N)^2}{np}\right)\right)$ & $\mathcal{O}\left(\frac{\lambda\xi\vartheta}{mpn}\right)$ & $\widetilde{\mathcal{O}}\left(\frac{\lambda\vartheta K_{PS}(\log K_{PS})^2}{mn}\right)$ \\ \hline
  DFT-P codes \cite{Gunduz20} & $\left(\widetilde{\mathcal{O}}\left(\frac{\lambda\xi N(\log N)^{2}}{mp}\right),\widetilde{\mathcal{O}}\left(\frac{\xi\vartheta N(\log N)^2}{np}\right)\right)$ & $\mathcal{O}\left(\frac{\lambda\xi\vartheta}{mpn}\right)$ & $\widetilde{\mathcal{O}}\left(\frac{\lambda\vartheta mn(\log mn)^2}{mn}\right)$ \\ \hline
  E-EP codes \cite{EP code} & $\left(\widetilde{\mathcal{O}}\left(\frac{\lambda\xi N(\log N)^{2}}{mp}\right),\widetilde{\mathcal{O}}\left(\frac{\xi\vartheta N(\log N)^2}{np}\right)\right)$ & $\mathcal{O}\left(\frac{\lambda\xi\vartheta}{mpn}\right)$ & $\widetilde{\mathcal{O}}\left(\frac{\lambda\vartheta K_{EP}(\log K_{EP})^2}{mn}\right)$ \\ \hline
  Degraded Strategy-2 & $\left(\widetilde{\mathcal{O}}\left(\frac{\lambda\xi N(\log N)^{2}}{mp}\right),\widetilde{\mathcal{O}}\left(\frac{\xi\vartheta N(\log N)^2}{np}\right)\right)$ & $\mathcal{O}\left(\frac{\lambda\xi\vartheta}{mpn}\right)$ & $\widetilde{\mathcal{O}}\left(\frac{\lambda\vartheta K_{D2}(\log K_{D2})^2}{mn}\right)$ \\ \hline
  \end{tabular}}
\begin{tablenotes}
       \footnotesize
       \item[]\quad\quad\quad\,\, Here, $K_{SD}=
      \left\{\begin{array}{@{}ll}
                                (m+\big\lceil\frac{X}{p}\big\rceil)p(n+1)+p\big\lceil\frac{X}{p}\big\rceil-1,&\text{if}~\big\lceil\frac{X}{p}\big\rceil=\frac{X}{p},p<m \\
                                (m+\big\lceil\frac{X}{p}\big\rceil)p(n+1)-p\big\lceil\frac{X}{p}\big\rceil+2X-1,&\text{if}~\big\lceil\frac{X}{p}\big\rceil>\frac{X}{p},p<m \\
                                m(pn+n\big\lceil\frac{X}{\min\{m,n\}}\big\rceil-\big\lceil\frac{X}{\min\{m,n\}}\big\rceil)+mp+2X-1, &\text{if}~p\geq m
                                \end{array}
      \right.,$
      \item[] \quad\quad\quad\,\, $R(m,p,n)$ denotes the bilinear complexity for multiplying two matrices with sizes of $m$-by-$p$ and
      \item[] \quad\quad\quad\,\,\, $p$-by-$n$ \cite{Smirnov,Strassen}, and $K_{D2}=\min\{(m+1)(np+X)-1,(n+1)(mp+X)-1\}$.
     \end{tablenotes}
\end{table*}

For the case of arbitrary partition from Table \ref{tab:comparative456}, an elementary calculation shows that
\begin{enumerate}
\item In terms of recovery threshold, download cost and decoding complexity, our degraded strategy-2 is superior to S-GPD codes for all parameters $m,p,n,X$ and  PS codes for $X<\max\{mp,np\}$, while upload cost and complexity for encoding and server computation achieve the same performance.
\item Compared to DFT-P codes, our strategy achieves lower recover threshold and download cost when $X\geq\min\big\{\frac{np-1}{2mn-m-1},\frac{mp-1}{2mn-n-1} \big\}$, and the performance with respect to upload cost and complexity for encoding and server computation are the same. However, DFT-P codes require a lower decoding complexity since it is constructed based on discrete Fourier transform of finite field.
\item 
    To the best of our knowledge, the bilinear complexity $R(m,p,n)$ for matrix multiplication has not been determined  in its general form from the open literature.
Here, we compare them in Table \ref{tab:comparative22} for some special parameters. It is straightforward from Table \ref{tab:comparative22} to observe that our degraded strategy-2 achieves less recovery threshold when $X$ is less than some parameter threshold.
\end{enumerate}

\begin{table*}[htbp]
\centering
\caption{{Comparison of our Degraded Strategy-2 with E-EP codes for some parameters}}
\resizebox{165mm}{15mm}{
  \begin{tabular}{|c|c|c|c|c|}
  \hline
  \multirow{2}*{\minitab[c]{Partitioning \\ Parameters}} & \multirow{2}*{\minitab[c]{Current Best Bilinear \\ Complexity $R$}}  & \multicolumn{2}{c|}{Recovery Threshold} &  \multirow{2}*{\minitab[c]{Improved \\ Regimes}}  \\ \cline{3-4}
   &  & E-EP codes $(K_{EP}=2R+2X-1)$ & Degraded Strategy-2 $(K_{D2})$ & \\ \hline
  $m=2,p=2,n=2$ & $7$ \cite{Strassen} & $13+2X$ & $3X+11$ & $X<2$ \\ \hline
  $m=3,p=3,n=3$ & $23$ \cite{Johnson} & $45+2X$ & $4X+35$ & $X<5$  \\ \hline
  $m=5,p=5,n=5$ & $98$ \cite{Sedoglavic} & $195+2X$ & $6X+149$ & $X<12$ \\ \hline
  $m=7,p=7,n=7$ & $250$ \cite{Sedoglavic} & $499+2X$ & $8X+391$ & $X<18$  \\\hline
  $m=9,p=9,n=9$ & $511$ \cite{Sedoglavic} & $1021+2X$ & $10X+809$ & $X<27$ \\\hline
  \end{tabular}}
  \label{tab:comparative22}
\end{table*}



The problem of SMBMM was first introduced in \cite{Chen and Jafar} under the special case $X=X_{\mathbf{A}}=X_{\mathbf{B}}$, which is compared in Table \ref{tab:comparative33} with our strategy.
The recovery threshold, download cost and decoding complexity of our strategy are shown to outperform \cite{Chen and Jafar} when $X\leq\frac{\max\{np,mp\}}{G+1}$, so does the amount of common randomness when $X<\max\{\frac{(n-1)p}{G+1},\frac{mpn-2mp+p}{(G+1)(n-1)}\}$. It is also worthy pointing out that, this advantage is obtained without any additional penalty on upload cost, encoding complexity and server computation complexity.

\begin{table*}[htbp]\centering
\caption{{Comparison for secure multi-party batch matrix multiplication}}\label{tab:comparative33}
\resizebox{165mm}{20mm}{
  \begin{tabular}{|@{}c@{}|@{}c@{}|@{}c@{}|@{}c@{}|}
  \hline
  SMBMM & Our Degraded Strategy & Chen \emph{et al.} Strategy \cite{Chen and Jafar} & Improved Regimes \\ \hline
  Recovery Threshold $K$ & $K_{SM}$ & $K_C=(LG+L)mpn+2X-1$ &  $X\leq\frac{\max\{np,mp\}}{G+1}$    \\
  \hline
  Amount of Common Randomness $\rho$ & $\frac{K_{SM}}{GLmn}-1$  & $\frac{(GL+L)mpn-mp+p+X-1}{GLmn}-1$ & $X<\max\{\frac{(n-1)p}{G+1},\frac{mpn-2mp+p}{(G+1)(n-1)}\}$  \\ \hline
  Upload Cost of Sources $(U_{\mathbf{A}},U_{\mathbf{B}})$ & $\left(\frac{N}{Lmp},\frac{N}{Lnp}\right)$ & $\left(\frac{N}{Lmp},\frac{N}{Lnp}\right)$ & Equal \\ \hline
  Download Cost $D$ &  $\frac{K_{SM}}{GLmn}$ & $\frac{K_C}{GLmn}$ & $X\leq\frac{\max\{np,mp\}}{G+1}$   \\ \hline
  Encoding Complexity $(\mathcal{C}_{\mathbf{A}},\mathcal{C}_{\mathbf{B}})$ & $\left(\widetilde{\mathcal{O}}\left(\frac{\lambda\xi N\log^2N}{Lmp}\right),\widetilde{\mathcal{O}}\left(\frac{\xi\vartheta N\log^2N}{Lnp}\right)\right)$ & $\left(\widetilde{\mathcal{O}}\left(\frac{\lambda\xi N\log^2N}{Lmp}\right),\widetilde{\mathcal{O}}\left(\frac{\xi\vartheta N\log^2N}{Lnp}\right)\right)$ & Equal \\ \hline
  Server Computation Complexity $\mathcal{C}_s$ & $\mathcal{O}\left(\frac{\lambda\xi\vartheta}{Lmpn}\right)$ & $\mathcal{O}\left(\frac{\lambda\xi\vartheta}{Lmpn}\right)$ & Equal \\ \hline
  Decoding Complexity $\mathcal{C}_d$ & $\widetilde{\mathcal{O}}\left(\frac{\lambda\vartheta K_{SM}\log^2K_{SM}}{LGmn}\right)$ & $\widetilde{\mathcal{O}}\left(\frac{\lambda\vartheta K_C\log^2K_C}{LGmn}\right)$ & $X\leq\frac{\max\{np,mp\}}{G+1}$ \\ \hline
  \end{tabular}}
\begin{tablenotes}
       \footnotesize
       \item[]Here, $K_{SM}=\min\{(LG+L-1)mpn+np+(Gm-G+m)X-1, (LG+L-1)mpn+mp+(Gn-G+n)X-1\}$.
\end{tablenotes}
\end{table*}



\section{Conclusion}\label{conclusion}
In this paper, we first introduced  a novel coded computation strategy for SSMM, then it was extended to the more general SMBMM setup. The extended strategy focuses on the tradeoff between recovery threshold, system cost and complexity, based on the idea of cross subspace alignment that facilitates a form of interference alignment of Cauchy-Vandermonde structure to align the desired terms in server responses along the Cauchy part and align the interference along dimensions corresponding to the Vandermonde part. Notably, in the degraded case considered by Chen \emph{et al.} \cite{Chen and Jafar} that the two batches of data matrices have same security level (i.e., $X=X_{\mathbf{A}}=X_{\mathbf{B}}$), the proposed strategy for SMBMM has advantage in recovery threshold, download cost, decoding complexity and the amount of common randomness, without any additional penalty with respect to upload cost and  complexity for encoding and server computation, when $X$ is smaller than some threshold.
It is also valuable to point out that, exploring the relationships between the securities of data matrices for the servers and the user, and jointly designing the randomness to protect such securities  are interesting research directions in the future.
\begin{appendix}[Proof of Lemma \ref{lemma:prod}]
In fact, for any fixed group $h\in[G]$, by \eqref{encoding function:h,1:1}-\eqref{security:multi-party2}, 
\begin{IEEEeqnarray}{rCl}
\widetilde{\mathbf{A}}^{h}(\alpha)\cdot\widetilde{\mathbf{B}}^{h}(\alpha)
&=& \frac{\prod_{\ell=2}^{L}(f_{h,\ell}-\alpha)^{\kappa}}{(f_{h,1}-\alpha)^{\psi}}\sum\limits_{r=0}^{\delta}\mathbf{H}_r^{h,1}(f_{h,1}-\alpha)^r\notag\\
&&+\sum\limits_{\ell=2}^{L}\frac{(f_{h,1}-\alpha)^{\psi}\cdot\prod_{k\in[2:L]\backslash\{\ell\}}(f_{h,k}-\alpha)^{\kappa}}{(f_{h,\ell}-\alpha)^{\kappa}}
\sum\limits_{r=0}^{\kappa+p-2}\mathbf{H}_r^{h,\ell}(f_{h,\ell}-\alpha)^r \notag\\
&&+\sum\limits_{\ell=2}^{L}\Bigg(\prod_{k\in[2:L]\backslash\{\ell\}}(f_{h,k}-\alpha)^{\kappa}\Bigg)\Big(\mathbf{P}^{h,1}(\alpha)\cdot\mathbf{Q}^{h,\ell}(\alpha)+\mathbf{P}^{h,\ell}(\alpha)\cdot\mathbf{Q}^{h,1}(\alpha)\Big)\notag\\
&&+\sum\limits_{s\in[2:L]}\sum\limits_{t\in[2:L]\backslash\{s\}}\Bigg((f_{h,1}-\alpha)^{\psi}\cdot\prod_{\ell\in[2:L]\backslash\{s,t\}}(f_{h,\ell}-\alpha)^{\kappa}\Bigg)\mathbf{P}^{h,s}(\alpha)\cdot\mathbf{Q}^{h,t}(\alpha) \notag\\
&=&\frac{\prod_{\ell=2}^{L}(f_{h,\ell}-\alpha)^{\kappa}}{(f_{h,1}-\alpha)^{\psi}}\sum\limits_{r=0}^{\psi-1}\mathbf{H}_r^{h,1}(f_{h,1}-\alpha)^{r}\notag\\
&&+\sum\limits_{\ell=2}^{L}\frac{(f_{h,1}-\alpha)^{\psi}\cdot\prod_{k\in[2:L]\backslash\{\ell\}}(f_{h,k}-\alpha)^{\kappa}}{(f_{h,\ell}-\alpha)^{\kappa}}\sum\limits_{r=0}^{\kappa-1}\mathbf{H}_r^{h,\ell}(f_{h,\ell}-\alpha)^{r}+  \Gamma^{h}(\alpha),\IEEEeqnarraynumspace \label{expanding form}
\end{IEEEeqnarray}
where
\begin{subequations}
\begin{IEEEeqnarray}{rCl}
\Gamma^{h}(\alpha)&\triangleq&\left(\prod_{\ell=2}^{L}(f_{h,\ell}-\alpha)^{\kappa}\right)\sum\limits_{r=\psi}^{\delta}\mathbf{H}_r^{h,1}(f_{h,1}-\alpha)^{r-\psi}\label{eqn:degree:1}\\
&&+\sum\limits_{\ell=2}^{L}\left((f_{h,1}-\alpha)^{\psi}\cdot\prod_{k\in[2:L]\backslash\{\ell\}}(f_{h,k}-\alpha)^{\kappa}\right)
\sum\limits_{r=\kappa}^{\kappa+p-2}\mathbf{H}_r^{h,\ell}(f_{h,\ell}-\alpha)^{r-\kappa}\\
&&+\sum\limits_{\ell=2}^{L}\Bigg(\prod_{k\in[2:L]\backslash\{\ell\}}(f_{h,k}-\alpha)^{\kappa}\Bigg)\mathbf{P}^{h,1}(\alpha)\mathbf{Q}^{h,\ell}(\alpha)\label{eqn:degree}\\
&&+\sum\limits_{\ell=2}^{L}\Bigg(\prod_{k\in[2:L]\backslash\{\ell\}}(f_{h,k}-\alpha)^{\kappa}\Bigg)\mathbf{P}^{h,\ell}(\alpha)\mathbf{Q}^{h,1}(\alpha)\\
&&+\sum\limits_{s\in[2:L]}\sum\limits_{t\in[2:L]\backslash\{s\}}\Bigg((f_{h,1}-\alpha)^{\psi}\cdot\prod_{\ell\in[2:L]\backslash\{s,t\}}(f_{h,\ell}-\alpha)^{\kappa}\Bigg)\mathbf{P}^{h,s}(\alpha)\mathbf{Q}^{h,t}(\alpha).\label{eqn:degree:2}
\end{IEEEeqnarray}
\end{subequations}

The first term in \eqref{expanding form} is
\begin{IEEEeqnarray}{rCl}
&&\frac{\prod_{\ell=2}^{L}(f_{h,\ell}-\alpha)^{\kappa}}{(f_{h,1}-\alpha)^{\psi}}\sum\limits_{r=0}^{\psi-1}\mathbf{H}_r^{h,1}(f_{h,1}-\alpha)^{r}\notag\\
&\overset{(a)}{=}&\left(\sum\limits_{s=0}^{\psi-1}c_{s}^{h,1}\cdot(f_{h,1}-\alpha)^{s-\psi}+\sum\limits_{s=\psi}^{(L-1)\kappa}c_{s}^{h,1}\cdot(f_{h,1}-\alpha)^{s-\psi}\right) \sum\limits_{r=0}^{\psi-1}\mathbf{H}_r^{h,1}(f_{h,1}-\alpha)^{r}\label{example_pSDMM122} \notag\\
&{=}&\left(\sum\limits_{s=1}^{\psi}\frac{c_{\psi-s}^{h,1}}{(f_{h,1}-\alpha)^{s}}+\sum\limits_{s=\psi}^{(L-1)\kappa}c_{s}^{h,1}\cdot(f_{h,1}-\alpha)^{s-\psi}\right) \sum\limits_{r=0}^{\psi-1}\mathbf{H}_r^{h,1}(f_{h,1}-\alpha)^{r}\notag\\
&=&\sum\limits_{r=0}^{\psi-1}\sum\limits_{s=r+1}^{\psi}\frac{c_{\psi-s}^{h,1}\mathbf{H}_r^{h,1}}{(f_{h,1}-\alpha)^{s-r}} \notag\\
&&+\underbrace{\sum\limits_{r=0}^{\psi-1}\sum\limits_{s=1}^{r}c_{\psi-s}^{h,1}\mathbf{H}_r^{h,1}(f_{h,1}-\alpha)^{r-s}+
\sum\limits_{s=\psi}^{(L-1)\kappa}c_{s}^{h,1}(f_{h,1}-\alpha)^{s-\psi}\sum\limits_{r=0}^{\psi-1}\mathbf{H}_r^{h,1}(f_{h,1}-\alpha)^{r}}_{\triangleq\Gamma^{h,1}(\alpha)},\notag
\end{IEEEeqnarray}
where $(a)$ follows by expanding the numerator polynomial $\prod_{\ell=2}^{L}\big((f_{h,1}-\alpha)+(f_{h,\ell}-f_{h,1})\big)^{\kappa}$ as $\prod_{\ell=2}^{L}\big((f_{h,1}-\alpha)+(f_{h,\ell}-f_{h,1})\big)^{\kappa}=\sum\limits_{s=0}^{(L-1)\kappa}c_s^{h,1}(f_{h,1}-\alpha)^{s}$
with $c_s^{h,1}$ being the coefficients of $(f_{h,1}-\alpha)^{s}$, and by setting $c_{s}^{h,1}=0$ 
for any $s\in[(L-1)\kappa+1:\psi-1]$ if $(L-1)\kappa<\psi-1$. Notice that we use the convention $\sum_{s=\psi}^{(L-1)\kappa}c_{s}^{h,1}\cdot(f_{h,1}-\alpha)^{s-\psi}=0$ if $(L-1)\kappa<\psi$.


Similarly, the second term in \eqref{expanding form} has
\begin{IEEEeqnarray}{c}
\frac{(f_{h,1}-\alpha)^{\psi}\cdot\prod_{k\in[2:L]\backslash\{\ell\}}(f_{h,k}-\alpha)^{\kappa}}{(f_{h,\ell}-\alpha)^{\kappa}}\sum\limits_{r=0}^{\kappa-1}\mathbf{H}_r^{h,\ell}(f_{h,\ell}-\alpha)^{r}
= \sum\limits_{r=0}^{\kappa-1}\sum\limits_{s=r+1}^{\kappa}\frac{c_{\kappa-s}^{h,\ell}}{(f_{h,\ell}-\alpha)^{s-r}}\mathbf{H}_r^{h,\ell} \notag\\ \quad\quad\quad\quad\quad+\underbrace{\sum\limits_{r=0}^{\kappa-1}\sum\limits_{s=1}^{r}c_{\kappa-s}^{h,\ell}\cdot\mathbf{H}_r^{h,\ell}(f_{h,\ell}-\alpha)^{r-s}
\sum\limits_{s=\kappa}^{(L-2)\kappa+\psi}c_{s}^{h,\ell}\cdot(f_{h,\ell}-\alpha)^{s-\kappa}\sum\limits_{r=0}^{\kappa-1}\mathbf{H}_r^{h,\ell}(f_{h,\ell}-\alpha)^{r}}_{\triangleq\Gamma^{h,\ell}(\alpha)},\notag
\end{IEEEeqnarray}
where $c_s^{h,\ell}$ is the coefficients of $(f_{h,\ell}-\alpha)^{s}$ in the numerator polynomial $\big((f_{h,\ell}-\alpha)+(f_{h,1}-f_{h,\ell})\big)^{\psi}\cdot\prod_{k\in[2:L]\backslash\{\ell\}}\big((f_{h,\ell}-\alpha)+(f_{h,k}-f_{h,\ell})\big)^{\kappa}$, i.e.,
\begin{IEEEeqnarray*}{c}
\big((f_{h,\ell}-\alpha)+(f_{h,1}-f_{h,\ell})\big)^{\psi}\cdot\prod_{k\in[2:L]\backslash\{\ell\}}\big((f_{h,\ell}-\alpha)+(f_{h,k}-f_{h,\ell})\big)^{\kappa}=\sum\limits_{s=0}^{(L-2)\kappa+\psi}c_s^{h,\ell}(f_{h,\ell}-\alpha)^{s}.
\end{IEEEeqnarray*}


It is easy to prove that the degree of $\Gamma^h(\alpha)$ is determined by the term in \eqref{eqn:degree} when $m>1$ since its degree is the largest over the terms in \eqref{eqn:degree:1}-\eqref{eqn:degree:2},
given by $\deg\big(\Gamma^{h}(\alpha)\big)=(L-1)mnp+np+X_{\mathbf{A}}+(m-1)X_{\mathbf{B}}-2$.
  Moreover, the degrees of $\Gamma^{h,\ell}(\alpha)$ ($\ell\in[L]$) are given by $ \deg\big(\Gamma^{h,1}(\alpha)\big)=(L-1)mnp-1$, and $\deg\big(\Gamma^{h,\ell}(\alpha)\big)=(L-1)mnp+(m-1)X_{\mathbf{B}}-1$ for $\ell\in[2:L]$.
Therefore, $\max\big\{\deg\big(\Gamma^{h}(\alpha)\big),\deg\big(\Gamma^{h,1}(\alpha)\big),\ldots,\deg\big(\Gamma^{h,L}(\alpha)\big)\big\}$ is given by $ (L-1)mnp+np+X_{\mathbf{A}}+(m-1)X_{\mathbf{B}}-2=\phi$, where $\phi$ is defined in \eqref{definition:phi}.

Then, the product $\widetilde{\mathbf{A}}^{h}(\alpha)\cdot\widetilde{\mathbf{B}}^{h}(\alpha)$ can be written as:
\begin{IEEEeqnarray*}{rCl}
\widetilde{\mathbf{A}}^{h}(\alpha)\cdot\widetilde{\mathbf{B}}^{h}(\alpha)
&=&\sum\limits_{r=0}^{\psi-1}\sum\limits_{s=r+1}^{\psi}\frac{c_{\psi-s}^{h,1}\mathbf{H}_r^{h,1}}{(f_{h,1}-\alpha)^{s-r}}+\sum\limits_{\ell=2}^{L}\sum\limits_{r=0}^{\kappa-1}\sum\limits_{s=r+1}^{\kappa}\frac{c_{\kappa-s}^{h,\ell}\mathbf{H}_r^{h,\ell}}{(f_{h,\ell}-\alpha)^{s-r}}+
\Gamma^{h}(\alpha)+\Gamma^{h,1}(\alpha)+\sum\limits_{\ell=2}^{L}\Gamma^{h,\ell}(\alpha)\notag\\
&=&\sum\limits_{r=0}^{\psi-1}\sum\limits_{s=r+1}^{\psi}\frac{c_{\psi-s}^{h,1}\mathbf{H}_r^{h,1}}{(f_{h,1}-\alpha)^{s-r}}+\sum\limits_{\ell=2}^{L}\sum\limits_{r=0}^{\kappa-1}\sum\limits_{s=r+1}^{\kappa}\frac{c_{\kappa-s}^{h,\ell}\mathbf{H}_r^{h,\ell}}{(f_{h,\ell}-\alpha)^{s-r}}+\sum\limits_{r=0}^{\phi}\mathbf{U}_{r}^{h}\alpha^{r},
\end{IEEEeqnarray*}
where $\mathbf{U}_{r}^{h}$ is the interference corresponding to the term $\alpha^r$ for any $r\in[0:\phi]$, which represents various combinations of sub-matrix products and can be found explicitly by expanding $\Gamma^{h}(\alpha)+\Gamma^{h,1}(\alpha)+\sum_{\ell=2}^{L}\Gamma^{h,\ell}(\alpha)$ but the exact form is unimportant.

\end{appendix}


\begin{thebibliography}{99}


\bibitem{GSPolyDot code} M. Aliasgari, O. Simeone, and J. Kliewer, ``Private and secure distributed matrix multiplication with flexible communication load,'' \textit{IEEE Trans. Inf. Forensics Security}, vol. 15, pp. 2722-2734, 2020.

%

\bibitem{Tandon secure code} W.-T. Chang and R. Tandon, ``On the capacity of secure distributed matrix multiplication,'' \textit{2018 IEEE Global Communications Conference (GLOBECOM)}, Abu Dhabi, United Arab Emirates, 2018, pp. 1-6.

\bibitem{Chen and Jafar} Z. Chen, Z. Jia, Z. Wang, and S.A. Jafar, ``GCSA codes with noise alignment for secure coded multi-party batch matrix multiplication,'' 
\emph{IEEE J. Selt. Area. Inf. Theory,} vol. 2, no. 1, pp. 306-316, March 2021.

\bibitem{Tail1} J. Dean and L.A. Barroso, ``The tail at scale,'' \textit{Communications of the ACM}, vol. 56, no. 2, pp. 74-80, 2013.

\bibitem{MapReduce} J. Dean and S. Ghemawat, ``MapReduce: simplified data processing on large clusters,'' \textit{Sixth USENIX Symposium on Operating System Design and Implementation}, Dec. 2004.


{\bibitem{Rouayheb secure code} R.G.L. D'Oliveira, S.E. Rouayheb, and D. Karpuk, ``GASP codes for secure distributed matrix multiplication,'' \textit{IEEE Trans. Inf. Theory}, vol. 66, no. 7, pp. 4038-4050, July 2020.}


\bibitem{GpolyDot} S. Dutta, Z. Bai, H. Jeong, T.M. Low, and P. Grover, ``A unified coded deep neural network training strategy based on generalized polydot codes for matrix multiplication.'' [Online]. Available: https://arxiv.org/abs/1811.10751, 2018.

\bibitem{Dutta} S. Dutta, V. Cadambe, and P. Grover, ``Short-dot: computing large linear transforms distributedly using coded short dot products,'' in \textit{Proc. 29th Annu. Conf. Neural Inf. Process. Syst. (NIPS)}, Barcelona, Spain, Dec. 2016, pp. 2100-2108.


\bibitem{MatDot code} S. Dutta, M. Fahim, F. Haddadpour, H. Jeong, V. Cadambe, and P. Grover, ``On the optimal recovery threshold of coded matrix multiplication,'' \textit{IEEE Trans. Inf. Theory}, vol. 66, no. 1, pp. 278-301, Jan. 2020.





\bibitem{CSA_PIR} Z. Jia, H. Sun, and S. A. Jafar, ``Cross subspace alignment and the asymptotic capacity of $X$-secure $T$-private information retrieval,'' \textit{IEEE Trans. Inf. Theory}, vol. 65, no. 9, pp. 5783-5798, Sept. 2019.

\bibitem{Jia and Jafar} Z. Jia and S.A. Jafar, ``Cross subspace alignment codes for coded distributed batch computation,'' to appear in \textit{IEEE Trans. Inf. Theory}, 2021.

\bibitem{Johnson} R.W. Johnson and A.M. McLoughlin, ``Noncommutative bilinear algorithms for $3\times 3$ matrix multiplication,'' \textit{SIAM J. Comput.} 15, 595-603 (1986).

\bibitem{Kakar secure code}  J. Kakar, S. Ebadifar, and A. Sezgin, ``On the capacity and straggler-robustness of distributed secure matrix multiplication,'' \textit{IEEE Access}, vol. 7, pp. 45783-45799, Apr. 2019.

{\bibitem{Kakar and Khristoforov} J. Kakar, A. Khristoforov, S. Ebadifar, and A. Sezgin, ``Uplink-downlink tradeoff in secure distributed matrix multiplication.'' [Online]. Available: https://arxiv.org/abs/1910.13849, 2019.}

\bibitem{recommender} Y. Koren, R. Bell, and C. Volinsky, ``Matrix factorization techniques for recommender systems,'' \textit{Computer}, vol. 42, no. 8, pp. 30-37, 2009.


%

\bibitem{Lee} K. Lee, M. Lam, R. Pedarsani, D. Papailiopoulos, and K. Ramchandran, ``Speeding up distributed machine learning using codes,'' \textit{IEEE Trans. Inf. Theory}, vol. 64, pp. 1514-1529, March 2018.

\bibitem{Lee2} K. Lee, C. Suh, and K. Ramchandran, ``High-dimensional coded matrix multiplication,'' in \textit{Proc. Int. Symp. Inf. Theory (ISIT)}, Aachen, Germany, Jun. 2017, pp. 2418-2422.



{\bibitem{Gunduz20} N. Mital, C. Ling, and D. Gunduz, ``Secure distributed matrix computation with discrete fourier transform.'' [Online]. Available: https://arxiv.org/abs/2007.03972, 2020.}


\bibitem{EP SMC} H.A. Nodehi, S.R.H. Najarkolaei, and M.A. Maddah-Ali, ``Entangled polynomial coding in limited-sharing multi-party computation,'' \textit{2018 IEEE Information Theory Workshop (ITW)}, Guangzhou, 2018, pp. 1-5.

\bibitem{Limited-sharing22} H.A. Nodehi and M.A. Maddah-Ali, ``Secure coded multi-party computation for massive matrix operations,'' to appear in \textit{IEEE Trans. Inf. Theory}, 2021. 




\bibitem{Cauchy-Vandermonde matrix} O. Pordavi, ``Recent research on pure and applied algebra'', Nova Science Publishers, Inc., \textit{New York}, 2003.

\bibitem{Sedoglavic} A. Sedoglavic, Fast matrix multiplication algorithms, [Online]. Available: https://fmm.univ-lille.fr/



\bibitem{Shamir} A. Shamir, ``How to share a secret,'' \textit{Communications of the ACM}, vol. 22, pp. 612-613, 1979.

\bibitem{Smirnov} A.V. Smirnov, ``The bilinear complexity and practical algorithms for matrix multiplication,'' \textit{Computational Mathematics and Mathematical Physics}, vol. 53, pp. 1781-1795, Dec 2013.

\bibitem{Strassen} V. Strassen, ``Gaussian elimination is not optimal,'' \textit{Numer. Math.}, 13 (1969), pp. 354-356.



\bibitem{Von} J. Von Zur Gathen and J. Gerhard, \textit{Modern computer algebra.} Cambridge university press, 2013.

\bibitem{repetition} D. Wang, G. Joshi, and G. Wornell, ``Using straggler replication to reduce latency in large-scale parallel computing,'' \textit{ACM SIGMETRICS Perform. Eval. Rev.}, vol. 43, no. 3, pp. 7-11, Dec. 2015.


\bibitem{Tail3} N.J. Yadwadkar, B. Hariharan, J.E. Gonzalez, and R. Katz, ``Multi-task learning for straggler avoiding predictive job scheduling,'' \textit{Journal of Machine Learning Research}, vol. 17, no. 106, pp. 1-37, 2016.

\bibitem{Yang secure code} H. Yang and J. Lee, ``Secure distributed computing with straggling servers using polynomial codes,'' \textit{IEEE Trans. Inf. Forensics Security}, vol. 14, no. 1, pp. 141-150, Jan. 2019.

\bibitem{Yao} A.C. Yao, ``Protocols for secure computations,'' \textit{23rd Annual Symposium on Foundations of Computer Science (sfcs 1982)}, Chicago, IL, USA, 1982, pp. 160-164.


\bibitem{Polynomial code} Q. Yu, M.A. Maddah-Ali, and S. Avestimehr, ``Polynomial codes: an optimal design for high-dimensional coded matrix multiplication,'' in \textit{Proc. 30th Annu. Conf. Neural Inf. Process. Syst. (NIPS)}, Long Beach, CA, USA, Dec. 2017, pp. 4403-4413.

\bibitem{EP code} Q. Yu, M.A. Maddah-Ali, and A.S. Avestimehr, ``Straggler mitigation in distributed matrix multiplication: fundamental limits and optimal coding,'' \textit{IEEE Trans. Inf. Theory}, vol. 66, no. 3, pp. 1920-1933, March 2020.

\bibitem{Qian Yu} Q. Yu and A.S. Avestimehr, ``Entangled polynomial codes for secure, private, and batch distributed matrix multiplication: breaking the ``cubic'' barrier.'' [Online]. Available: https://arxiv.org/abs/2001.05101v1, 2020.

\bibitem{Spark} M. Zaharia, M. Chowdhury, M.J. Franklin, S. Shenker, and I. Stoica, ``Spark: cluster computing with working sets,'' in \textit{Proceedings of the 2nd USENIX HotCloud}, vol. 10, p. 10, June 2010.

\bibitem{Zhao} W. Zhao, X. Ming, S. Mikael, and P.H. Vincent, ``Secure degrees of freedom of wireless $X$ networks using artificial noise alignment,'' \textit{IEEE Transactions on communications}, vol. 63, no. 7, pp. 2632-2646, 2015.

\bibitem{batch matrix} J. Zhu and X. Tang, ``Secure batch matrix multiplication from grouping lagrange encoding,'' \textit{IEEE Communications Letters}, vol. 25, no. 4, pp. 1119-1123, April 2021.




\end{thebibliography}

\end{document}